\documentclass{aastex631}

\usepackage[T1]{fontenc}%
\usepackage{bm}
\usepackage{tabularx}
\usepackage{booktabs}
\usepackage{longtable,array,threeparttablex}
\usepackage{threeparttable}
\usepackage{ragged2e} 
\usepackage{booktabs,makecell, multirow, tabularx}

\usepackage{appendix}

\begin{document}

\title{The investigation of 84 TESS totally eclipsing contact binaries}

\author[0009-0008-3792-6444]{Yani Guo}
\affiliation{Shandong Key Laboratory of Optical Astronomy and Solar-Terrestrial Environment, School of Space Science and Physics, Institute of Space Sciences, Shandong University, Weihai, Shandong, 264209, China}

\author[0000-0003-3590-335X]{Kai Li}
\affiliation{Shandong Key Laboratory of Optical Astronomy and Solar-Terrestrial Environment, School of Space Science and Physics, Institute of Space Sciences, Shandong University, Weihai, Shandong, 264209, China}

\author[0009-0005-0485-418X]{Liheng Wang}
\affiliation{Shandong Key Laboratory of Optical Astronomy and Solar-Terrestrial Environment, School of Space Science and Physics, Institute of Space Sciences, Shandong University, Weihai, Shandong, 264209, China}

\author{Qiqi Xia}
\affiliation{Shandong Key Laboratory of Optical Astronomy and Solar-Terrestrial Environment, School of Space Science and Physics, Institute of Space Sciences, Shandong University, Weihai, Shandong, 264209, China}

\author[0009-0009-6364-0391]{Xiang Gao}
\affiliation{Shandong Key Laboratory of Optical Astronomy and Solar-Terrestrial Environment, School of Space Science and Physics, Institute of Space Sciences, Shandong University, Weihai, Shandong, 264209, China}

\author[0000-0003-2477-3430]{Jingran Xu}
\affiliation{Shandong Key Laboratory of Optical Astronomy and Solar-Terrestrial Environment, School of Space Science and Physics, Institute of Space Sciences, Shandong University, Weihai, Shandong, 264209, China}

\author{Jingyi Wang}
\affiliation{Shandong Key Laboratory of Optical Astronomy and Solar-Terrestrial Environment, School of Space Science and Physics, Institute of Space Sciences, Shandong University, Weihai, Shandong, 264209, China}

\correspondingauthor{Kai Li}
\email{kaili@sdu.edu.cn}

%% Note that the \and command from previous versions of AASTeX is now
%% depreciated in this version as it is no longer necessary. AASTeX 
%% automatically takes care of all commas and "and"s between authors names.

%% AASTeX 6.31 has the new \collaboration and \nocollaboration commands to
%% provide the collaboration status of a group of authors. These commands 
%% can be used either before or after the list of corresponding authors. The
%% argument for \collaboration is the collaboration identifier. Authors are
%% encouraged to surround collaboration identifiers with ()s. The 
%% \nocollaboration command takes no argument and exists to indicate that
%% the nearby authors are not part of surrounding collaborations.

%% Mark off the abstract in the ``abstract'' environment. 
\begin{abstract}

Based on the eclipsing binary catalog provided by \cite{2022ApJS..258...16P}, 84 totally eclipsing contact binaries with stable light curves were selected. The TESS light curves of these 84 targets were studied using the Physics Of Eclipsing Binaries code. The results indicate that there are 18 deep contact binaries, 39 moderate contact binaries, and 27 shallow contact binaries. Among them, 43 targets exhibit the O'Connell effect, which is attributed to the presence of star-spot on the component's surface. 15 targets are low-mass ratio deep contact binaries and may be contact binary merging candidates. Based on the relationship between the period and semi-major axis of contact binaries, their absolute physical parameters such as mass, radius, and luminosity were derived. The evolutionary status of these 84 targets was studied using the mass-luminosity and mass-radius relation diagrams. Their initial masses were also estimated. Our results are compared with those of targets that have been historically studied. 
Among the 84 targets, 44 targets have been studied before, and 21 of these have mass ratios $q$ that are consistent with historical values within a 10\% difference. For the inconsistent targets, we conducted a detailed investigation and found that the main reasons are poor quality of historical data, or the fact that the machine learning methods used in historical studies might not accurately determine the physical parameters for individual targets.

\end{abstract}

%% Keywords should appear after the \end{abstract} command. 
%% The AAS Journals now uses Unified Astronomy Thesaurus concepts:
%% https://astrothesaurus.org
%% You will be asked to selected these concepts during the submission process
%% but this old "keyword" functionality is maintained in case authors want
%% to include these concepts in their preprints.

%% From the front matter, we move on to the body of the paper.
%% Sections are demarcated by \section and \subsection, respectively.
%% Observe the use of the LaTeX \label
%% command after the \subsection to give a symbolic KEY to the
%% subsection for cross-referencing in a \ref command.
%% You can use LaTeX's \ref and \label commands to keep track of
%% cross-references to sections, equations, tables, and figures.
%% That way, if you change the order of any elements, LaTeX will
%% automatically renumber them.
%%
%% We recommend that authors also use the natbib \citep
%% and \citet commands to identify citations.  The citations are
%% tied to the reference list via symbolic KEYs. The KEY corresponds
%% to the KEY in the \bibitem in the reference list below. 

\section{Introduction} \label{sec:intro}

Contact binaries belong to one class of  close binary stars. When both components of a close binary star are overfilling with their Roche lobes, a contact binary is formed. The components of a contact binary are sharing a common envelope, mass transfer and energy exchange can be achieved between the two components through the common envelope. With the material exchange between the two components, many physical phenomena occurs, such as O’Connell effect \citep{1951PRCO....2...85O,2003ChJAA...3..142L,2009PASP..121.1366L,2013NewA...22...57L,2014ApJS..212....4Q,2014NewA...30...64L,2015NewA...41...17L,2016AJ....151...67Z},  the minimum mass ratio \citep{1995ApJ...444L..41R, 2006MNRAS.369.2001L, 2007MNRAS.377.1635A, 2010MNRAS.405.2485J, 2019PASP..131e4203C, 2021ApJ...922..122L, 2022AJ....164..202L} and so on. 

Due to the influence of various complex factors on the evolution process of contact binary stars, the complete evolutionary theory  and model remain controversial and require further research. Some researchers proposed that contact binary stars are evolved from short period detached binary stars through angular momentum loss via magnetic braking \citep{1988ASIC..241..345G, 1994ASPC...56..228B, 2002ApJ...575..461E, 2003MNRAS.342.1260Q, 2006AcA....56..347S, 2017RAA....17...87Q} or nuclear evolution \citep{2014MNRAS.437..185Y}, and later will evolve into  rapidly rotating single stars such as blue stragglers or FK Com stars \citep{1995ApJ...444L..41R, 2005AJ....130.1206Q, 2007MNRAS.377.1635A, 2010MNRAS.405.2485J, 2016RAA....16...68Z}.  Researchers have proposed many hypotheses regarding the details of the evolution of contact binaries merging into single stars.  For example, \cite{2006Ap&SS.304...25Q} proposed that deep (the fillout factor f>50\%), low-mass ratio (q<0.25) contact binaries are at the late stage of contact binary evolution. As the fillout factor increases, the contact system becomes more dynamically unstable and is expected to merge into a rapidly rotating single star \citep{1995ApJ...438..887R}.
Researchers have proposed that when the spin angular momentum of the binary system exceeds one-third of the orbital angular momentum, $J_{s}/J_{o}>1/3$, the binary star becomes dynamically unstable, and the mass ratio reaches the limit of the mass ratio \citep{1980A&A....92..167H,1995ApJ...438..887R, 2006MNRAS.369.2001L, 2007MNRAS.377.1635A, 2009MNRAS.394..501A, 2021MNRAS.501..229W, 2023MNRAS.519.5760L, 2024SerAJ.208....1A}.

Contact binary stars are divided into A-subtype contact binary stars and W-subtype ones \citep{1970VA.....12..217B}. A-subtype contact binary stars refer to the massive stars with high temperatures, while the W-subtype ones are exactly the opposite. According to \cite{1994ApJ...434..277W, 1995AJ....110..782W} , A-subtype contact binaries are in different phases of thermal relaxation oscillations. However, \cite{1985MNRAS.217..843M} argue that A-subtype contact binaries evolve from W-subtype binaries. The evolutionary pathways of the two subtypes remain controversial \citep{2020MNRAS.492.4112Z}.

O'Connell effect \citep{1951PRCO....2...85O} refers to the unequal heights of two maxima of light curves of the contact binary stars.  There are multiple explanations for the O'Connell effect, such as the star-spot model, since contact binaries are generally composed of two late-type components, which are usually associated with magnetic activity in their surfaces. Other explanations include hot spots formed by mass accretion \citep{1990BAAS...22.1296S} and circumstellar material \citep{2003ChJAA...3..142L}. The star-spot model is considered to be the most likely explanation (e.g., \citealt{2007ApJ...671..811Q, 2013AJ....145..100L, 2016NewA...47....3Z, 2016Ap&SS.361...63L, 2019MNRAS.485.4588L, 2024PASP..136b4201P}). 

% The determination of the mass ratio is very important for contact binary stars. Through radial velocity observations, accurate mass ratios can be obtained. However, most contact binaries lack radial velocity observations, so their mass ratios are typically determined using the q-search method based on photometric light curves, which is widely used and accepted (e.g., \citealt{2006AJ....131.3028Q, 2015MNRAS.448.2890D, 2019MNRAS.485.4588L, 2020AJ....159..189L,2024NewA..11102252Y}). Researches have confirmed that the mass ratios obtained from the totally eclipsing light curves of contact binary stars through photometric studies are reliable and consistent with those obtained from spectroscopic observations \citep{2003CoSka..33...38P, 2005Ap&SS.296..221T, 2021AJ....162...13L}.   

Due to the development of large-scale survey projects, a large amount of data is provided for us to study the physical parameters of contact binary stars, such as the Transiting Exoplanet Survey Satellite (TESS; \citealt{2015JATIS...1a4003R}), the Gaia mission \citep{2018A&A...616A...1G}, Guoshoujing Telescope (the Large Sky Area Multi-Object Fiber Spectroscopic Telescope, LAMOST) \citep{2015RAA....15.1095L} , Kepler mission \citep{2010Sci...327..977B, 2010ApJ...713L..79K}, the Zwicky Transient Facility survey (ZTF; \citealt{2020ApJS..249...18C} ) and the Catalina Sky Survey (CSS; \citealt{2017MNRAS.465.4678M}). 
This paper selected TESS data, which provides a large amount of high-precision, high-time-resolution continuous light curves, to obtain accurate physical parameters for a large sample of contact binaries.

\begin{table}
\centering
\caption{Information of the 84 Targets.}
% \begin{threeparttable 
\label{tab:basic information}
% \scalebox{0.85}{
\begin{tabular}{ccccccccccccccc}
\hline
\hline
Targets(TIC) & R.A.  & Dec.  &  Period(d)  &  Sector  &  $T_{TESS}(K)$  & $T_{Gaia}(K)$ & $T_{LAMOST}(K)$  & $T_{mean}(K)$\\
\hline

101070833  &  03 49 10.60   &   -42 36 55.08  &   0.827664       &      1                                   &  6723          &      ...           &     ...    &      6720 \\   
114500611  &  00 47 16.04   &   -19 41 43.71  &   0.488774       &      3                                   &  6809          &      6588          &     ...    &      6700 \\   
115105140  &  21 05 58.99   &   -36 15 33.99  &   0.410686       &      1                                   &  6229          &      6279          &     ...    &      6250 \\   
117978580  &  04 46 59.01   &   -14 37 23.30  &   0.527263       &      5                                   &  7134          &      ...           &     ...    &      7130 \\   
121134517  &  03 52 00.21   &   -21 55 48.35  &   0.335171       &      4                                   &  6269          &      6325          &     ...    &      6300 \\   
123038007  &  18 36 28.12   &   +45 56 23.18  &   0.422688       &      26                                  &  6774          &      6702          &     ...    &      6740 \\   
124205037  &  13 07 57.72   &   -31 00 48.06  &   0.332734       &      10                                  &  5572          &      5774          &     ...    &      5670 \\   
13070701   &  05 03 33.60   &   -25 21 55.77  &   0.414076       &      5                                   &  6797          &      6690          &     ...    &      6740 \\   
138256097  &  14 43 05.36   &   +30 24 35.50  &   0.463932       &      23-24,50                            &  6378          &      6704          &    6536    &      6540 \\   
140757590  &  04 57 07.48   &   -72 07 54.85  &   0.418404       &      9-13                                &  6386          &      6222          &     ...    &      6300 \\   
141126559  &  05 22 13.47   &   -71 56 18.67  &   0.776660       &      4-6                                 &  8389          &      9933          &     ...    &      9160 \\   
142587827  &  11 11 28.90   &    73 06 55.01  &   0.439398       &      20-21,40,47-48                      &  6696          &      ...           &     ...    &      6700 \\   
146520491  &  05 06 17.66   &   -20 07 52.79  &   0.448645       &      5, 32                               &  6478          &      6330          &     ...    &      6400 \\   
147083089  &  21 24 47.14   &   -47 10 50.45  &   0.362744       &      1                                   &  5547          &      5795          &     ...    &      5670 \\   
149837469  &  23 28 01.09   &   -33 59 51.61  &   0.385592       &      2                                   &  5578          &      6078          &     ...    &      5830 \\   
159939533  &  16 01 21.96   &    48 29 38.00  &   0.282255       &      24-25                               &  5068          &      5437          &    5089    &      5250 \\   
164720673  &  01 42 54.80   &   -20 07 28.16  &   0.365991       &      3                                   &  6211          &      ...           &     ...    &      6210 \\   
167673754  &  05 19 54.10   &   -35 54 06.34  &   0.565193       &      5                                   &  7410          &      ...           &     ...    &      7410 \\   
193580427  &  17 43 23.08   &   +47 51 41.79  &   0.394304       &      25-26                               &  6159          &      6096          &    6159    &      6140 \\   
193635719  &  17 47 43.87   &   +46 32 31.40  &   0.486618       &      25-26                               &  6920          &      6942          &     ...    &      6930 \\   
193823999  &  17 55 35.84   &    43 48 20.16  &   0.369873       &      25-26                               &  6233          &      6564          &    6237    &      6400 \\  
\hline
\end{tabular}
%\begin{tablenotes}[lr] %添加此处
%\footnotesize 
%         \item Here D23 refer to the values from \cite{2023MNRAS.525.4596D}, $diff=|This paper-D23|/This %paper$
%\end{tablenotes} %添加此处
%\end{threeparttable} %添加此处
\begin{tablenotes}  %增加注脚用，“[]”是左对齐的
\item \textbf{Note.}This table is available in its entirety in machine-readable form. 
\end{tablenotes} %增加注脚用
% \end{threeparttable} %增加注脚用
\end{table}

\section{Observations} \label{sec:style}

\subsection{Targets selection}
TESS, launched by NASA in April 2018, provides a large amount of high-quality photometric data. TESS's four wide-field cameras continuously monitor over 200,000 stars, capturing detailed light curves with different cadences, each observed for approximately 27 days. The 2-minute cadence data from sector 1 to sector 58 have been selected for analysis, which allows us to obtain accurate photometric parameters.

\cite{2022ApJS..258...16P} identified 4584 eclipsing binary stars from the 2-minute cadence data of about 200,000 stars in sectors 1-26 during the first two years of TESS, and presented a catalog website \footnote[1]{https://tessebs.villanova.edu/} containing locations, periods and basic light curve properties of these stars. Based on this catalog, we selected the target eclipsing binary stars for analysis.

Four criteria were used to select targets. The first criterion was that the targets were selected from the 4584 eclipsing binaries based on the morphological parameter proposed for classification from a dimensionality reduction algorithm \citep{2012AJ....143..123M, 2022ApJS..258...16P}. The morphology of eclipsing binary stars is a continuous variable within 0-1 defined according to the shape of the light curves. The closer the parameter is to 0, the greater the degree of separation of the binary stars. Conversely, the closer the parameter is to 1, the closer the system is to an ellipsoidal variable. Therefore, we selected the targets beyond 0.5 for analysis. The second criterion, the selected targets were cross-matched with Gaia DR3 \citep{2021A&A...649A...1G} by setting the match radius to 21 arcsec. We selected targets for which the total flux of other stars within 21 arcsec is less than one percent of the target star's flux. The third criterion, the totally eclipsing contact binary stars were selected through visual inspection according to their shapes of the light curves. The variability of the light curve over time generally results in changes to physical parameters of contact binaries. This paper aims to obtain precise physical parameters for each target, disregarding their temporal variations. Thus, the fourth criterion was that the targets whose light curves do not change with time were selected from the totally eclipsing contact binaries. Finally, 84 targets were selected for further analysis. The details of these targets are shown in Table \ref{tab:basic information}.

\begin{figure*}
 \includegraphics[width=.95\linewidth]{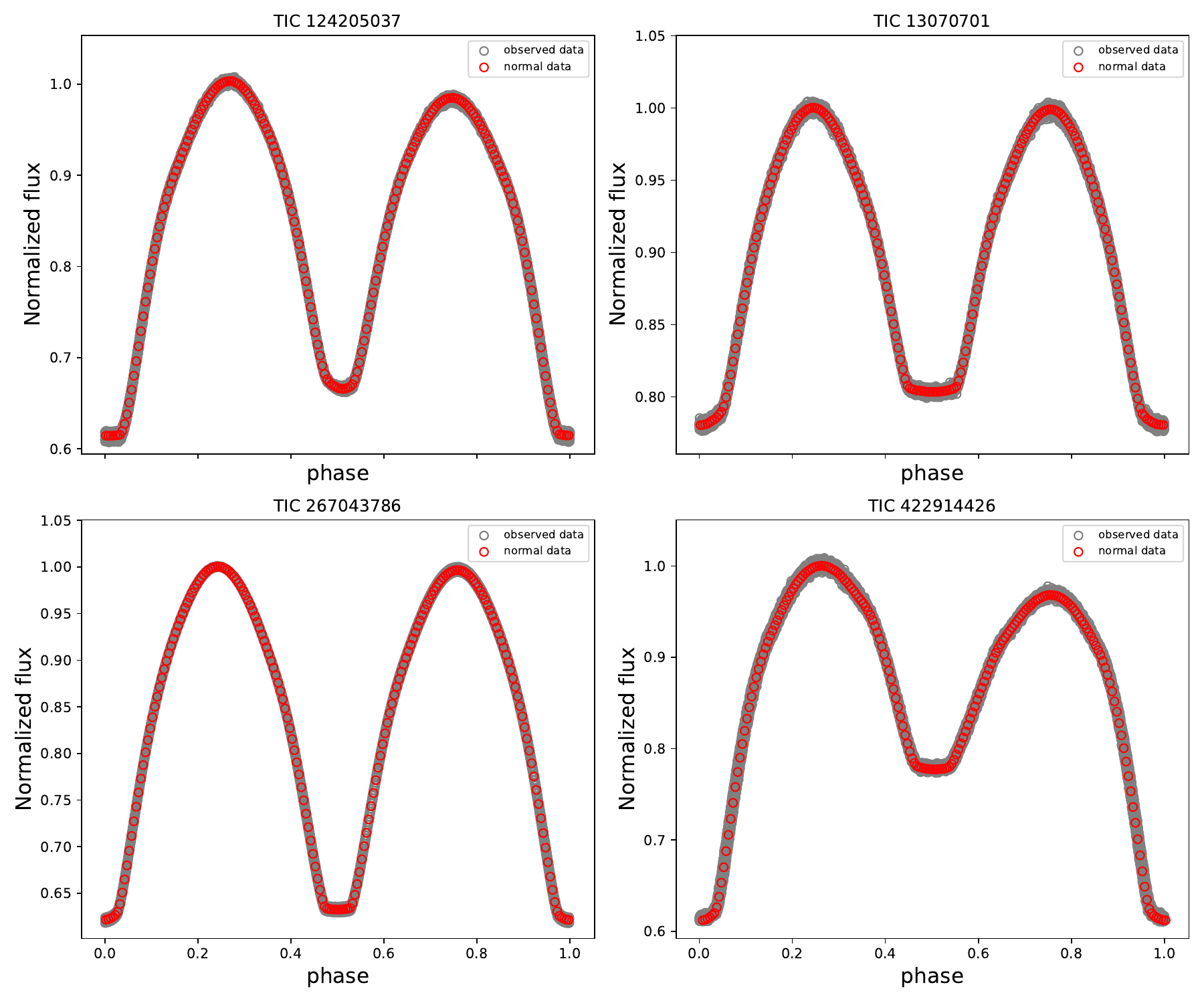}
 \caption{Four examples of normalized phase-folded light curves: the gray points represent observed data, while the red points represent normal data.}
 \label{bin-phase}
\end{figure*}

\subsection{Download and process data}
After completing the target selection, \emph{lightkurve} package \citep{2018ascl.soft12013L} was used to download data from the Mikulski Archive for Space Telescopes (MAST) archive. In order to obtain more precise physical parameters, the 2-minute cadence data were employed. The Pre Data Search Conditioning Simple Aperture Photometry (PDCSAP) light curves of TESS from sector 1 to sector 58 were primarily used, while the Simple Aperture Photometry (SAP) light curves were employed for cases where the PDCSAP data were inadequate. The total time span ranges from BJD 2458325 to BJD 2459910. Some targets have been observed by many sectors, but some targets have only one, so the specific time span varies. 

To save time for subsequent analysis, all the light curves were combined into 200 normal points calculated by averaging the light curves in phase. The time-series light curves were folded to phased light curves according to periods from the catalog of \cite{2022ApJS..258...16P}. The phased light curves with normalized flux are shown in Figure \ref{bin-phase}.

\section{photometric solutions}

\begin{table}
\centering
\caption{The photometric elements of the 84 contact binaries.}
% \begin{threeparttable 
\label{tab:solution}
% \scalebox{0.85}{
\begin{tabular}{ccccccccccccccc}
\hline
\hline
Targets(TIC) & $q$  & $T_{1}^{'}(K)$  &  $T_{2}^{'}(K)$  & $i(deg)$ &  $f$  &  $L_{1}$/$L_{T}$  &$r_{1}$ & $r_{2}$    \\
\hline

101070833 	   &  $0.320^{+0.002}_{-0.002} $     &   $7016\pm204 $   &  $ 5553\pm161  $     &  $82.3^{+0.3}_{-0.4} $     &     $  0.04^{+0.01}_{-0.00} $  &  $0.842^{+0.022}_{-0.025} $     &  $0.484^{+0.002}_{-0.002} $     &  $0.287^{+0.002}_{-0.001} $    \\ 
114500611 	   &  $0.155^{+0.002}_{-0.002} $     &   $6796\pm86  $   &  $ 6115\pm77   $     &  $84.1^{+1.3}_{-0.8} $     &     $  0.17^{+0.03}_{-0.03} $  &  $0.884^{+0.008}_{-0.009} $     &  $0.551^{+0.002}_{-0.003} $     &  $0.241^{+0.002}_{-0.002} $ \\    
115105140 	   &  $0.168^{+0.002}_{-0.002} $     &   $6260\pm63  $   &  $ 6202\pm63   $     &  $77.0^{+0.4}_{-0.4} $     &     $  0.30^{+0.02}_{-0.02} $  &  $0.814^{+0.009}_{-0.009} $     &  $0.550^{+0.002}_{-0.002} $     &  $0.253^{+0.003}_{-0.002} $    \\ 
117978580 	   &  $0.327^{+0.005}_{-0.004} $     &   $7225\pm185 $   &  $ 6861\pm176  $     &  $79.6^{+0.3}_{-0.3} $     &     $  0.31^{+0.02}_{-0.01} $  &  $0.723^{+0.027}_{-0.029} $     &  $0.498^{+0.002}_{-0.003} $     &  $0.307^{+0.003}_{-0.003} $ \\    
121134517 	   &  $0.185^{+0.001}_{-0.001} $     &   $6285\pm87  $   &  $ 6363\pm88   $     &  $77.0^{+0.2}_{-0.2} $     &     $  0.29^{+0.01}_{-0.01} $  &  $0.788^{+0.013}_{-0.014} $     &  $0.542^{+0.002}_{-0.001} $     &  $0.260^{+0.002}_{-0.002} $    \\ 
123038007 	   &  $0.147^{+0.001}_{-0.001} $     &   $6753\pm122 $   &  $ 6676\pm120  $     &  $89.2^{+4.1}_{-2.3} $     &     $  0.64^{+0.02}_{-0.02} $  &  $0.825^{+0.014}_{-0.016} $     &  $0.573^{+0.002}_{-0.002} $     &  $0.258^{+0.002}_{-0.002} $ \\    
124205037 	   &  $0.213^{+0.005}_{-0.002} $     &   $4871\pm94  $   &  $ 5499\pm106  $     &  $78.9^{+0.5}_{-0.4} $     &     $  0.36^{+0.03}_{-0.04} $  &  $0.318^{+0.017}_{-0.019} $     &  $0.274^{+0.003}_{-0.003} $     &  $0.534^{+0.002}_{-0.002} $ \\    
13070701       &  $0.103^{+0.001}_{-0.001} $     &   $6763\pm83  $   &  $ 6575\pm81   $     &  $78.6^{+0.3}_{-0.3} $     &     $  0.43^{+0.02}_{-0.02} $  &  $0.885^{+0.007}_{-0.008} $     &  $0.590^{+0.002}_{-0.002} $     &  $0.223^{+0.002}_{-0.002} $   \\  
138256097 	   &  $0.232^{+0.003}_{-0.002} $     &   $6560\pm63  $   &  $ 6467\pm62   $     &  $78.4^{+0.3}_{-0.3} $     &     $  0.30^{+0.02}_{-0.02} $  &  $0.764^{+0.010}_{-0.010} $     &  $0.525^{+0.002}_{-0.002} $     &  $0.278^{+0.002}_{-0.002} $ \\    
140757590 	   &  $0.188^{+0.002}_{-0.002} $     &   $6312\pm77  $   &  $ 6252\pm77   $     &  $85.2^{+1.5}_{-0.9} $     &     $  0.56^{+0.03}_{-0.03} $  &  $0.789^{+0.012}_{-0.012} $     &  $0.553^{+0.002}_{-0.002} $     &  $0.274^{+0.003}_{-0.003} $ \\

\hline
\end{tabular}
%\begin{tablenotes}[lr] %添加此处
%\footnotesize 
%         \item Here D23 refer to the values from \cite{2023MNRAS.525.4596D}, $diff=|This paper-D23|/This %paper$
%\end{tablenotes} %添加此处
%\end{threeparttable} %添加此处
\begin{tablenotes}  %增加注脚用，“[]”是左对齐的
\item \textbf{Note.}This table is available in its entirety in machine-readable form. 
\end{tablenotes} %增加注脚用
\end{table}

The Physics Of Eclipsing Binaries (PHOEBE) version 2.4.11 \citep{2016ApJS..227...29P} with Markov Chain Monte Carlo (MCMC) \citep{2013PASP..125..306F} method was used to analyze light curves.
Before using this software, we should first determine the temperatures of the primary stars. Here, we used the temperatures from Gaia , TESS , and LAMOST surveys, and took their average values as the final temperatures of the primary stars, the detailed information is listed in Table \ref{tab:basic information}. The gravity-darkening coefficients $g_{1}$, $g_{2}$ and the bolometric albedos $A_{1}$, $A_{2}$ were determined according to \cite{1967ZA.....65...89L} and \cite{1969AcA....19..245R}, respectively. 
For the components of contact binaries with temperatures above and below 7200K, the gravity-darkening coefficients were set as 1.0 and 0.32, respectively \citep{1924MNRAS..84..665V, 1967ZA.....65...89L}, and the bolometric albedos were set to 1.0 and 0.5, respectively \citep{1969AcA....19..245R}.
The atmospheric model of \cite{2004A&A...419..725C} was adopted. The limb-darkening coefficients were used PHOEBE's limb-darkening table. 

To quickly find convergent solutions, we used the method proposed by \cite{2024ApJ...976..223W} to determine preliminary values of some of the physical parameters. In short, for systems with symmetric light curves, we used a parameter search method to provide preliminary values for the mass ratio ($q$), temperature of the secondary component ($T_{2}$), orbital inclination ($i$), luminosity of the primary star ($L_{1}$), and fillout factor ($f$). The parameter search method involves traversing an equally spaced grid in the parameter space to determine the best-fitting parameters. For systems with asymmetric light curves, we used a genetic algorithm \citep{1992SciAm.267a..66H} to determine the mass ratio ($q$), temperature of the secondary component ($T_{2}$), orbital inclination ($i$), luminosity of the primary star ($L_{1}$), fill-out factor ($f$), and star-spot parameters. The star-spot parameters include latitude ($\theta$), longitude ($\lambda$), angular radius ($r_{s}$), and its relative temperature ($T_{s}$). For most targets with asymmetric light curves, we added spot on the primary star, but for some targets whose fitted light curves are not good with one spot, we added a spot to the primary component and also added a spot to the secondary component.
To quickly obtain the star-spots parameters, the latitude ($\theta$) was fixed at 90 degrees. Considering the cases that the mass ratios of some targets are greater than 1, we set the phase shift to 0.5, recalculated the physical parameters using the same method, and chose the parameters with the smallest residual sum of squares of the fitting residuals as the initial parameters. 

\begin{figure*}
 \includegraphics[width=.95\linewidth]{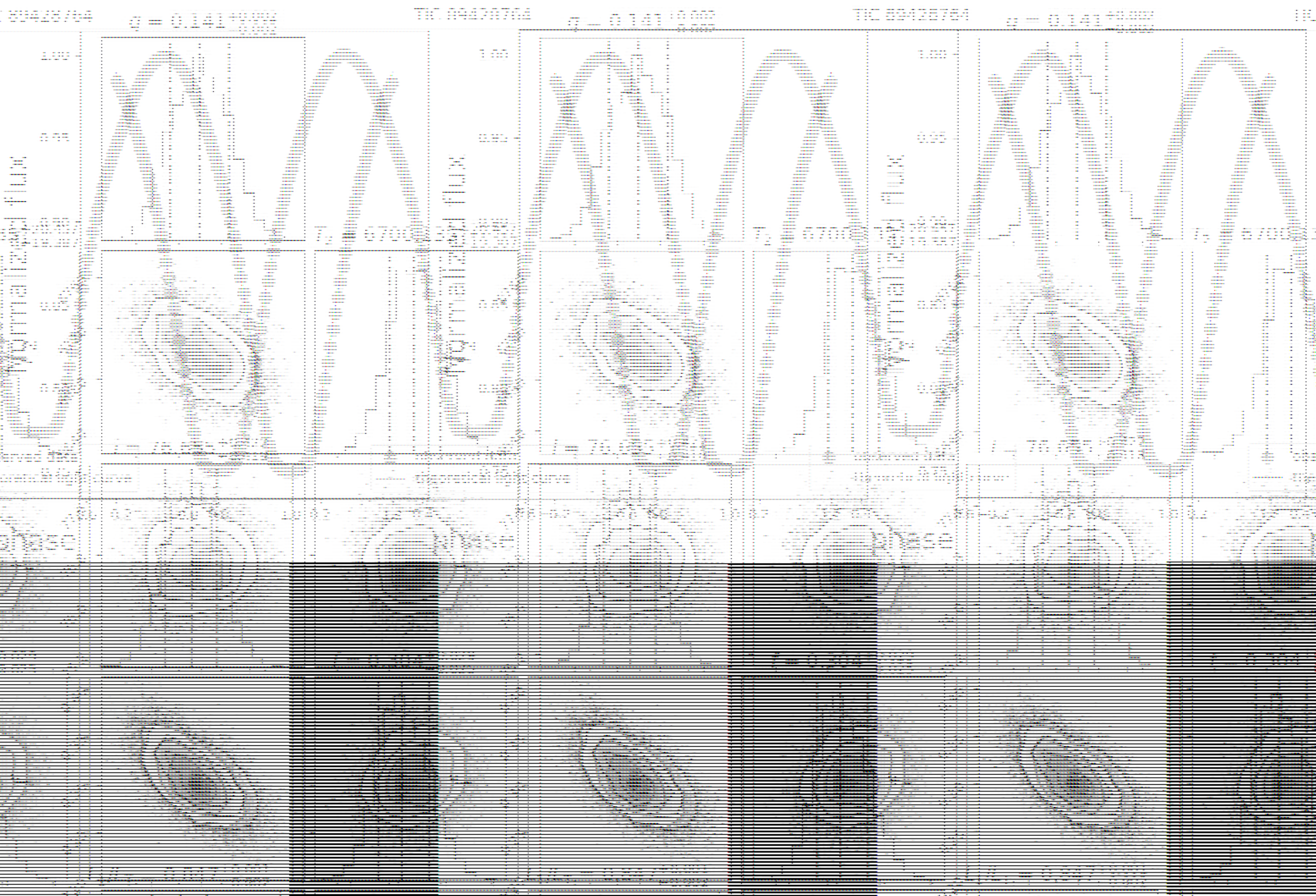}
 \caption{The posterior distributions of $q$, $T_{2}$, $i$, $f$, $L_{1}/L_{T}$ and fitting light curve of TIC 89428764. The fitting figures for all targets can be found in the online material of this journal.}
 \label{fitting1}
\end{figure*}

\begin{figure*}
 \includegraphics[width=.95\linewidth]{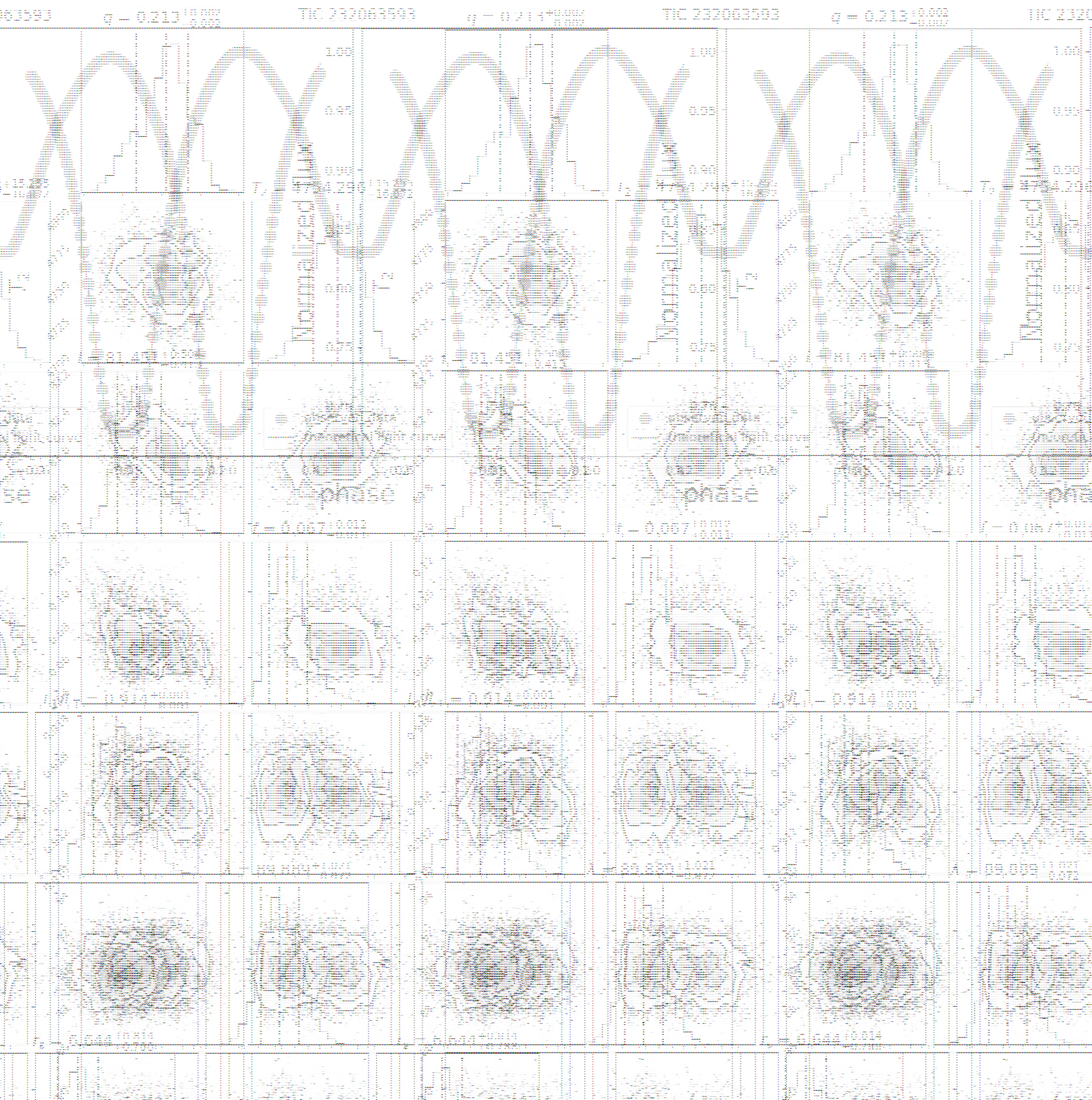}
 \caption{The posterior distributions of $q$, $T_{2}$, $i$, $f$, $L_{1}/L_{T}$, $\lambda$, $r_{s}$, $T_{s}$ and fitting light curve of TIC 232063593.}
 \label{fitting2}
\end{figure*}

Afterwards, the MCMC calculation was performed using a Gaussian distribution, with the obtained preliminary values serving as the mean of the Gaussian distribution. In this process, 20 walkers with 2000 iterations for each walker were adopted. 
The mass ratio ($q$), orbital inclination ($i$), fill-out factor ($f$), the temperature of the secondary star ($T_{2}$), the luminosity of primary star ($L_{1}$), longitude ($\lambda$), radius ($r_{s}$), and its relative temperature ($T_{s}$) can be determined by PHOEBE with MCMC sampler.
The posterior distributions of these physical parameters are shown in the left panels of Figures \ref{fitting1} and \ref{fitting2}, illustrated with two examples: one target without star-spot and one target with star-spot. The physical parameters are listed in Table \ref{tab:solution}. 
The mean temperatures from Gaia, TESS, and LAMOST surveys are the average values of the primary and secondary stars, so we corrected the temperatures of primary and secondary stars according to the equations from \cite{2003A&A...404..333Z} and \cite{2013AJ....146..157C}:
\begin{equation}
T_{1}^{'} = (((1+k^{2})T_{mean}^{4}/(1+k^{2}(T_{2}/T_{1})^{4}))^{0.25},
\end{equation}
\begin{equation}
T_{2}^{'} = T_{1}(T_{2}/T_{1})
\end{equation}
where $T_{2}/T_{1}$ and $k$ refer to temperature and radius ratio, respectively. The temperatures and radii of primary and secondary stars $T_{1}$, $T_{2}$, $r_{1}$, $r_{2}$ are obtained from PHOEBE with MCMC. The values of $T_{1}^{'}$ and $T_{2}^{'}$ are listed in Table \ref{tab:solution}. Given the uncertainties of $T_{1}$, these are propagated to $T_{2}$ via the aforementioned equation and subsequently to $L_{1}/L_{T}$ through PHOEBE with MCMC. Therefore, in Table \ref{tab:solution}, the uncertainties for all parameters other than these three are due to the fitting process.
The spot information is listed in Table \ref{tab:spot} in the appendix. The corresponding fitting light curves are presented in right panels of Figures \ref{fitting1} and \ref{fitting2}. 
From Figures \ref{fitting1} and \ref{fitting2}, the fitting curves show that the theoretical light curves fit the observational data well.
% The results of all the targets are listed in Table \ref{tab:solution}. The spot information is listed in Table \ref{tab:spot} in the appendix. 
% $q$, $T_{2}$, $i$, $f$, $L_{1}/L_{T}$ of
% The posterior distributions of these physical parameters are shown in the left panels of Figures \ref{fitting1} and \ref{fitting2}, illustrated with two examples: one target without star-spot and one target with star-spot. The corresponding fitting light curves are presented in right panels of Figures \ref{fitting1} and \ref{fitting2}. 
% From Figures \ref{fitting1} and \ref{fitting2}, the fitting curves show that the theoretical light curves fit the observational data well.
It should be declared that the mass ratio refers to the ratio of the less massive component to the more massive one, so all the mass ratios are less than 1. Hence, the primary stars refer to the more massive components, and the secondary stars refer to the less massive ones.

\begin{table}
\caption{Fundamental parameters of the 84 targets.}
\hspace*{-2cm} % 调整此值以移动表格
\scalebox{0.80}{
\label{tab:absolute parameters}
\begin{tabular}{cccccccccccc}
\hline
\hline
Targets(TIC) & $M_{1}$($M_{\odot}$)  & $M_{2}$($M_{\odot}$)  & $L_{1}$($L_{\odot}$)  & $L_{2}$($L_{\odot}$) & $R_{1}$($R_{\odot}$) & $R_{2}$($R_{\odot}$) & $a$($R_{\odot}$) & $log J_{orb}($cgs$) $ & $M_{1i}$($M_{\odot}$) & $M_{2i}$($M_{\odot}$) \\
\hline
    101070833 & $3.93^{+0.07}_{-0.06}$ & $0.646^{+0.011}_{-0.010}$ & $2.025^{+0.050}_{-0.049}$ & $0.647^{+0.016}_{-0.016}$ & $2.493^{+0.020}_{-0.020}$ & $1.480^{+0.012}_{-0.012}$ & $5.15\pm0.04$ & 52.041  & 1.827  & 1.235  \\
    114500611 & $1.86^{+0.04}_{-0.04}$ & $0.252^{+0.005}_{-0.005}$ & $1.695^{+0.052}_{-0.051}$ & $0.263^{+0.008}_{-0.008}$ & $1.801^{+0.018}_{-0.018}$ & $0.787^{+0.008}_{-0.008}$ & $3.27\pm0.03$ & 51.541  & 1.288  & 1.474  \\
    115105140 & $1.05^{+0.02}_{-0.02}$ & $0.213^{+0.005}_{-0.005}$ & $1.512^{+0.051}_{-0.050}$ & $0.254^{+0.009}_{-0.008}$ & $1.545^{+0.017}_{-0.017}$ & $0.710^{+0.008}_{-0.008}$ & $2.81\pm0.03$ & 51.467  & 1.120  & 1.421  \\
    117978580 & $2.06^{+0.04}_{-0.04}$ & $0.662^{+0.013}_{-0.013}$ & $1.542^{+0.046}_{-0.045}$ & $0.505^{+0.015}_{-0.015}$ & $1.737^{+0.017}_{-0.017}$ & $1.070^{+0.010}_{-0.010}$ & $3.49\pm0.03$ & 51.788  & 1.197  & 1.533  \\
    121134517 & $0.73^{+0.02}_{-0.02}$ & $0.172^{+0.004}_{-0.004}$ & $1.322^{+0.050}_{-0.049}$ & $0.245^{+0.009}_{-0.009}$ & $1.279^{+0.016}_{-0.016}$ & $0.612^{+0.008}_{-0.008}$ & $2.36\pm0.03$ & 51.380  & 0.947  & 1.359  \\
    123038007 & $1.52^{+0.03}_{-0.03}$ & $0.289^{+0.006}_{-0.006}$ & $1.566^{+0.052}_{-0.051}$ & $0.231^{+0.008}_{-0.008}$ & $1.650^{+0.018}_{-0.018}$ & $0.745^{+0.008}_{-0.008}$ & $2.88\pm0.03$ & 51.441  & 1.115  & 1.571  \\
    124205037 & $0.44^{+0.01}_{-0.01}$ & $0.070^{+0.002}_{-0.002}$ & $1.286^{+0.049}_{-0.048}$ & $0.274^{+0.010}_{-0.010}$ & $1.252^{+0.016}_{-0.010}$ & $0.642^{+0.008}_{-0.008}$ & $2.34\pm0.03$ & 51.417  & ...    & ...    \\
    13070701  & $1.57^{+0.04}_{-0.03}$ & $0.201^{+0.005}_{-0.004}$ & $1.609^{+0.054}_{-0.053}$ & $0.166^{+0.006}_{-0.005}$ & $1.671^{+0.019}_{-0.019}$ & $0.631^{+0.007}_{-0.007}$ & $2.83\pm0.03$ & 51.309  & 1.153  & 1.523  \\
    138256097 & $1.37^{+0.03}_{-0.03}$ & $0.363^{+0.008}_{-0.007}$ & $1.540^{+0.049}_{-0.048}$ & $0.358^{+0.011}_{-0.011}$ & $1.638^{+0.017}_{-0.017}$ & $0.868^{+0.009}_{-0.009}$ & $3.12\pm0.03$ & 51.630  & 1.163  & 1.480  \\
    140757590 & $1.12^{+0.03}_{-0.02}$ & $0.260^{+0.006}_{-0.006}$ & $1.503^{+0.050}_{-0.049}$ & $0.283^{+0.009}_{-0.009}$ & $1.578^{+0.017}_{-0.017}$ & $0.784^{+0.009}_{-0.009}$ & $2.86\pm0.03$ & 51.512  & 1.108  & 1.460  \\

\hline
\end{tabular}}
%\begin{tablenotes}[lr] %添加此处
%\footnotesize 
%         \item Here D23 refer to the values from \cite{2023MNRAS.525.4596D}, $diff=|This paper-D23|/This %paper$
%\end{tablenotes} %添加此处
%\end{threeparttable} %添加此处
\begin{tablenotes}[]  %增加注脚用，“[]”是左对齐的
\item \hspace{0.4cm} \textbf{Note.}This table is available in its entirety in machine-readable form. 
\end{tablenotes} %增加注脚用
\end{table}

\section{Discussion and Conclusions}

\begin{table}[h!]
\hspace*{-3.5cm} % 调整此值以移动表格
\scalebox{0.75}{
\begin{threeparttable}
\caption{Historical studies and Comparative result of some parameters with this paper.} 

%\contcaption{A table continued from the previous one.}
\label{tab:comparison}
% \hspace{-10cm}
\begin{tabular}{ccccccccccccccccc}
%\begin{tabular}{@{}lcccccccccccccccccc@{}}  %加“@{}”左对齐贴紧
\hline
\hline
\multicolumn{1}{c}{ \multirow{2}*{Targets(TIC)} }  & \multicolumn{3}{c}{$q$} &  \multicolumn{3}{c}{$i$} & \multicolumn{3}{c}{$T_{2}/T_{1}$} & \multicolumn{3}{c}{$L_{2}/L_{1}$ \tnote{c}}  & \multicolumn{3}{c}{$f$}  &\multirow{2}*{REF\tnote{d}} \\
%\cline{2-4}
\cmidrule(r){2-4} \cmidrule(r){5-7} \cmidrule(r){8-10}  \cmidrule(r){11-13}  \cmidrule(r){11-13}  \cmidrule(r){14-16}
& This paper  & History   &  diff\tnote{a} & This paper  & History   & diff & This paper  & History   & diff &  This paper  & History   & diff & This paper  & History   & diff \\
\hline
 114500611        &        0.155                &        0.177                &        0.13                &        84.1                &        89.4                &        0.06                &        0.90                &        0.87                &        0.03                &        0.131                        &        0.136                &        0.04                &            0.17          &       0.24         &    0.29    &   D23        \\
 117978580        &        0.327                &        0.347                &        0.06                &        79.6                &        80.9                &        0.02                &        0.95                &        0.98                &        0.03                &        0.382                        &        0.367                &        0.04                &            0.31          &       0.34         &    0.09    &   D23        \\
 123038007        &        0.147                &        0.168                &        0.12                &        89.2                &        86.9                &        0.03                &        0.99                &        0.97                &        0.02                &        0.213                        &        0.203                &        0.05                &            0.64          &       0.69         &    0.08    &   D23        \\
 13070701         &        0.103                &        0.119                &        0.13                &        78.6                &        79.2                &        0.01                &        0.97                &        0.97                &        0.00                &        0.131                        &        0.142                &        0.08                &            0.43          &       0.38         &    0.13    &   D23        \\
 138256097        &        0.232                &        0.252                &        0.08                &        78.4                &        80.0                &        0.02                &        0.99                &        0.98                &        0.01                &        0.308                        &        0.282                &        0.09                &            0.30          &       0.35         &    0.13    &   D23        \\
 140757590        &        0.188                &        0.217                &        0.13                &        85.2                &        88.9                &        0.04                &        0.99                &        0.98                &        0.01                &        0.267                        &        0.261                &        0.02                &            0.56          &       0.74         &    0.24    &   D23        \\
 142587827        &        0.147                &        0.177                &        0.17                &        82.0                &        86.9                &        0.06                &        0.99                &        0.97                &        0.02                &        0.213                        &        0.208                &        0.02                &            0.67          &       0.51         &    0.31    &   D23        \\
 146520491        &        0.136                &        0.153                &        0.11                &        74.5                &        76.0                &        0.02                &        0.98                &        0.97                &        0.01                &        0.188                        &        0.176                &        0.07                &            0.41          &       0.33         &    0.25    &   D23        \\
 147083089        &        0.434                &        0.453                &        0.04                &        86.1                &        85.9                &        0.00                &        1.05                &        1.02                &        0.03                &        0.614                        &        ...                  &        ...                 &            0.13          &       0.11         &    0.21    &   S16        \\
 149837469        &        0.401                &        0.439                &        0.09                &        83.4                &        84.2                &        0.01                &        1.04                &        1.04                &        0.00                &        0.561                        &        0.549                &        0.02                &            0.05          &       0.03         &    0.89    &   S17        \\
 159939533        &        0.364                &        0.370                &        0.01                &        89.8                &        88.0                &        0.02                &        1.06                &        1.05                &        0.01                &        0.593                        &        0.520                &        0.14                &            0.20          &       0.20         &    0.01    &   M23        \\
 164720673        &        0.127                &        0.142                &        0.11                &        78.1                &        78.9                &        0.01                &        1.01                &        1.01                &        0.00                &        0.190                        &        0.189                &        0.00                &            0.52          &       0.50         &    0.04    &   D23        \\
 193823999        &        0.214                &        0.247                &        0.13                &        86.3                &        88.7                &        0.03                &        1.01                &        0.99                &        0.02                &        0.299                        &        0.282                &        0.06                &            0.30          &       0.27         &    0.15    &   D23        \\
 199612934        &        0.094                &        0.112                &        0.16                &        82.1                &        85.5                &        0.04                &        0.92                &        0.91                &        0.01                &        0.086                        &        0.109                &        0.21                &            0.46          &       0.31         &    0.49    &   D23        \\
 206537272        &        0.269                &        0.288                &        0.07                &        81.6                &        79.4                &        0.03                &        1.05                &        0.99                &        0.06                &        0.420                        &        0.267                &        0.57                &            0.55          &       0.74         &    0.26    &   U15        \\
 207174531        &        0.401                &        0.470(s)\tnote{b}    &        0.15                &        85.3                &        ...                 &        0.00                &        0.77                &        ...                 &        0.00                &        0.219                        &        ...                  &        ...                 &            0.08          &       ...          &    ...     &   D07        \\
 219738202        &        0.280                &        0.290(s)\tnote{b}    &        0.03                &        80.0                &        ...                 &        0.00                &        0.95                &        0.97                &        0.03                &        0.337                        &        ...                  &        ...                 &            0.54          &       ...          &    ...     &   P04        \\
 241393039        &        0.372                &        0.466                &        0.20                &        82.8                &        89.5                &        0.07                &        1.03                &        0.98                &        0.05                &        0.517                        &        0.479                &        0.08                &            0.13          &       0.28         &    0.54    &   D23        \\
 257273424        &        0.329                &        0.219                &        0.50                &        80.2                &        80.2                &        0.00                &        1.02                &        0.97                &        0.05                &        0.464                        &        0.249                &        0.86                &            0.23          &       0.64         &    0.64    &   D23        \\
 261089147        &        0.238                &        0.215                &        0.11                &        79.8                &        79.5                &        0.00                &        0.90                &        0.88                &        0.02                &        0.240                        &        0.125                &        0.92                &            0.35          &       0.10         &    2.46    &   L80        \\
 267043786        &        0.301                &        0.354(s)\tnote{b}    &        0.15                &        82.3                &        76.2                &        0.08                &        0.98                &        0.98                &        0.00                &        0.379                        &        0.388                &        0.02                &            0.27          &       0.58         &    0.54    &   C01        \\
 268686130        &        0.170                &        0.188                &        0.09                &        82.2                &        83.8                &        0.02                &        0.96                &        0.96                &        0.00                &        0.203                        &        0.201                &        0.01                &            0.24          &       0.25         &    0.04    &   D23        \\
 269505148        &        0.224                &        0.249                &        0.10                &        79.9                &        80.2                &        0.00                &        0.84                &        0.73                &        0.15                &        0.159                        &        ...                  &        ...                 &            0.13          &       ...          &    ...     &   R00        \\
 289166536        &        0.241                &        0.266                &        0.09                &        84.6                &        87.7                &        0.03                &        1.00                &        0.99                &        0.01                &        0.341                        &        0.309                &        0.10                &            0.49          &       0.54         &    0.09    &   D23        \\
 290034099        &        0.301                &        0.306                &        0.02                &        80.7                &        81.8                &        0.01                &        0.94                &        0.97                &        0.03                &        0.355                        &        0.325                &        0.09                &            0.39          &       0.47         &    0.18    &   N02        \\
 29777922         &        0.095                &        0.113                &        0.16                &        78.9                &        78.7                &        0.00                &        0.88                &        0.86                &        0.02                &        0.075                        &        0.094                &        0.20                &            0.39          &       0.41         &    0.05    &   D23        \\
 314459000        &        0.176                &        0.200                &        0.12                &        79.2                &        81.1                &        0.02                &        1.02                &        1.02                &        0.00                &        0.285                        &        0.268                &        0.06                &            0.61          &       0.62         &    0.02    &   D23        \\
 323442776        &        0.283                &        0.304                &        0.07                &        78.8                &        80.1                &        0.02                &        0.99                &        0.98                &        0.01                &        0.374                        &        0.328                &        0.14                &            0.28          &       0.33         &    0.17    &   D23        \\
 377296926        &        0.167                &        0.179                &        0.07                &        75.7                &        76.6                &        0.01                &        0.98                &        0.98                &        0.00                &        0.242                        &        0.221                &        0.10                &            0.62          &       0.75         &    0.18    &   D23        \\
 392536812        &        0.088                &        0.099                &        0.11                &        75.4                &        76.3                &        0.01                &        1.00                &        0.99                &        0.01                &        0.125                        &        0.135                &        0.07                &            0.55          &       0.54         &    0.01    &   D23        \\
 393943031        &        0.152                &        0.173                &        0.12                &        80.2                &        82.2                &        0.02                &        0.99                &        0.98                &        0.01                &        0.209                        &        0.201                &        0.04                &            0.46          &       0.42         &    0.12    &   D23        \\
 415969184        &        0.139                &        0.170                &        0.18                &        95.8                &        88.0                &        0.09                &        0.90                &        0.84                &        0.07                &        0.129                        &        0.142                &        0.09                &            0.41          &       0.34         &    0.21    &   D23        \\
 43864479         &        0.180                &        0.198                &        0.09                &        82.7                &        83.5                &        0.01                &        0.99                &        0.98                &        0.01                &        0.252                        &        0.233                &        0.08                &            0.48          &       0.57         &    0.16    &   D23        \\
 441128066        &        0.342                &        0.345                &        0.01                &        81.9                &        81.1                &        0.01                &        0.99                &        0.99                &        0.00                &        0.440                        &        0.387                &        0.14                &            0.27          &       0.29         &    0.06    &   P24        \\
 445924091        &        0.084                &        0.105                &        0.20                &        74.1                &        77.1                &        0.04                &        1.01                &        0.99                &        0.02                &        0.126                        &        0.136                &        0.08                &            0.68          &       0.34         &    1.02    &   D23        \\
 455064763        &        0.341                &        0.330                &        0.03                &        90.1                &        89.9                &        0.00                &        1.05                &        1.04                &        0.01                &        0.523                        &        0.487                &        0.07                &            0.12          &       0.13         &    0.05    &   A21        \\
 53842685         &        0.090                &        0.080                &        0.12                &        78.4                &        76.2                &        0.03                &        0.99                &        0.99                &        0.00                &        0.119                        &        0.126                &        0.06                &            0.49          &       0.90         &    0.45    &   W06        \\
 55753802         &        0.201                &        0.206                &        0.02                &        78.9                &        78.7                &        0.00                &        0.99                &        1.00                &        0.01                &        0.288                        &        0.413                &        0.30                &            0.53          &       0.59         &    0.10    &   A14        \\
 56001210         &        0.227                &        0.251                &        0.10                &        89.8                &        89.3                &        0.01                &        1.01                &        1.01                &        0.00                &        0.320                        &        0.302                &        0.06                &            0.25          &       0.31         &    0.18    &   D23        \\
 67686223         &        0.445                &        0.546                &        0.19                &        88.1                &        89.8                &        0.02                &        0.93                &        0.94                &        0.01                &        0.410                        &        ...                  &        ...                 &            0.05          &       0.16         &    0.68    &   L96        \\
 75377304         &        0.177                &        0.173                &        0.02                &        76.4                &        78.2                &        0.02                &        0.99                &        1.01                &        0.02                &        0.227                        &        0.189                &        0.20                &            0.47          &       0.40         &    0.17    &   G16        \\
 89428764         &        0.141                &        0.147                &        0.04                &        76.7                &        76.9                &        0.00                &        0.97                &        0.96                &        0.01                &        0.181                        &        0.156                &        0.16                &            0.30          &       0.33         &    0.07    &   T22        \\
 8989778          &        0.194                &        0.192                &        0.01                &        75.9                &        76.4                &        0.01                &        0.90                &        0.86                &        0.04                &        0.195                        &        0.131                &        0.49                &            0.32          &       0.33         &    0.03    &   G22        \\
 93053299         &        0.128                &        0.147                &        0.13                &        77.3                &        78.6                &        0.02                &        0.99                &        0.98                &        0.01                &        0.181                        &        0.182                &        0.00                &            0.51          &       0.52         &    0.01    &   D23        \\
\hline
\end{tabular} 
%\begin{tablenotes}[lr] %添加此处
%\footnotesize 
%         \item Here D23 refer to the values from \cite{2023MNRAS.525.4596D}, $diff=|This paper-D23|/This %paper$
%\end{tablenotes} %添加此处
%\end{threeparttable} %添加此处
% \hspace{1cm} % 调整此值以移动表格
% \hspace{4cm}
\end{threeparttable}}
\tablecomments{\newline$^{a}$ $diff=|This$ $paper-History|/History$. \newline $^{b}$ $s$ refers to the spectroscopic mass ratio. \newline $^{c}$ For the targets with multiple bands, we took their average value. \newline $^{d}$ Here D23 refers to \cite{2023MNRAS.525.4596D}. L80 refers to \cite{1980AJ.....85.1098L}. R00 refers to \cite{2000PASP..112..123R}. T22 refers to \cite{2022RAA....22c5009T}. S16 refers to \cite{2016AJ....152..219S}. S17 refers to \cite{2017JAVSO..45....3S}. L96 refers to \cite{1996A&AS..118..453L}. D07 refers to \cite{2007AJ....133..169D}. U15 refers to \cite{2015NewA...41....1U}. P04 refers to \cite{2004AJ....127.1712P}. C01 refers to \cite{2001CoSka..31....5C}. N02 refers to \cite{2002IBVS.5285....1N}. W06 refers to \cite{2006Ap&SS.301..195W}. A14 refers to \cite{2014NewA...31....1A}. G16 refers to \cite{2016NewA...44...40G}. P24 refers to \cite{2024arXiv240518618P}.
}

% \item \hspace{4cm} \textbf{Note.} 
% \item[\hspace{4cm} 
% \item[\hspace{4cm} b]$s$ refers to the spectroscopic mass ratio.
% \item[\hspace{4cm} c]For the targets with multiple bands, we took their average value.
% \item[\hspace{4cm} d]Here D23 refers to \cite{2023MNRAS.525.4596D}. L80 refers to \cite{1980AJ.....85.1098L}. R00 refers to \cite{2000PASP..112..123R}. T22 refers to \cite{2022RAA....22c5009T}. S16 refers to \cite{2016AJ....152..219S}. 
% \item[\hspace{4cm}]S17 refers to \cite{2017JAVSO..45....3S}. L96 refers to \cite{1996A&AS..118..453L}. D07 refers to \cite{2007AJ....133..169D}. U15 refers to \cite{2015NewA...41....1U}. \item[\hspace{4cm}]P04 refers to \cite{2004AJ....127.1712P}. C01 refers to \cite{2001CoSka..31....5C}. N02 refers to \cite{2002IBVS.5285....1N}. W06 refers to \cite{2006Ap&SS.301..195W}. A14 refers to \cite{2014NewA...31....1A}. \item[\hspace{4cm}]G16 refers to \cite{2016NewA...44...40G}.P24 refers to \cite{2024arXiv240518618P}.

\end{table}

% \begin{tablenotes}[]  %增加注脚用，“[]”是左对齐的
% \item \hspace{4cm} \textbf{Note.} 
% \item[\hspace{4cm} a]$diff=|This$ $paper-History|/History$. 
% \item[\hspace{4cm} b]$s$ refers to the spectroscopic mass ratio.
% \item[\hspace{4cm} c]For the targets with multiple bands, we took their average value.
% \item[\hspace{4cm} d]Here D23 refers to \cite{2023MNRAS.525.4596D}. L80 refers to \cite{1980AJ.....85.1098L}. R00 refers to \cite{2000PASP..112..123R}. T22 refers to \cite{2022RAA....22c5009T}. S16 refers to \cite{2016AJ....152..219S}. 
% \item[\hspace{4cm}]S17 refers to \cite{2017JAVSO..45....3S}. L96 refers to \cite{1996A&AS..118..453L}. D07 refers to \cite{2007AJ....133..169D}. U15 refers to \cite{2015NewA...41....1U}. \item[\hspace{4cm}]P04 refers to \cite{2004AJ....127.1712P}. C01 refers to \cite{2001CoSka..31....5C}. N02 refers to \cite{2002IBVS.5285....1N}. W06 refers to \cite{2006Ap&SS.301..195W}. A14 refers to \cite{2014NewA...31....1A}. \item[\hspace{4cm}]G16 refers to \cite{2016NewA...44...40G}.P24 refers to \cite{2024arXiv240518618P}.
% \end{tablenotes} %增加注脚用
% \end{threeparttable}}

Through analyzing the light curves of 84 targets, it was found that there are 18 deep contact binaries, 39 moderate contact binaries and 27 shallow contact binaries, among which 15 targets are low-mass ratio deep contact binary stars. 67 targets are A-subtype contact binaries, while 17 targets are W-subtype contact binaries.
There are 43 targets exhibiting O’Connell effect, which is interpreted using the star-spot model. It is important to highlight that the spot model involves uncertainties and degeneracies and thus should be considered merely as a reference for one potential interpretation. Therefore, in our subsequent analysis, we have categorized the targets into two groups: one exhibiting the O'Connell effect in their light curves, and the other without this effect.

\begin{figure*}
 \includegraphics[width=0.8\textwidth, height=8cm, keepaspectratio]{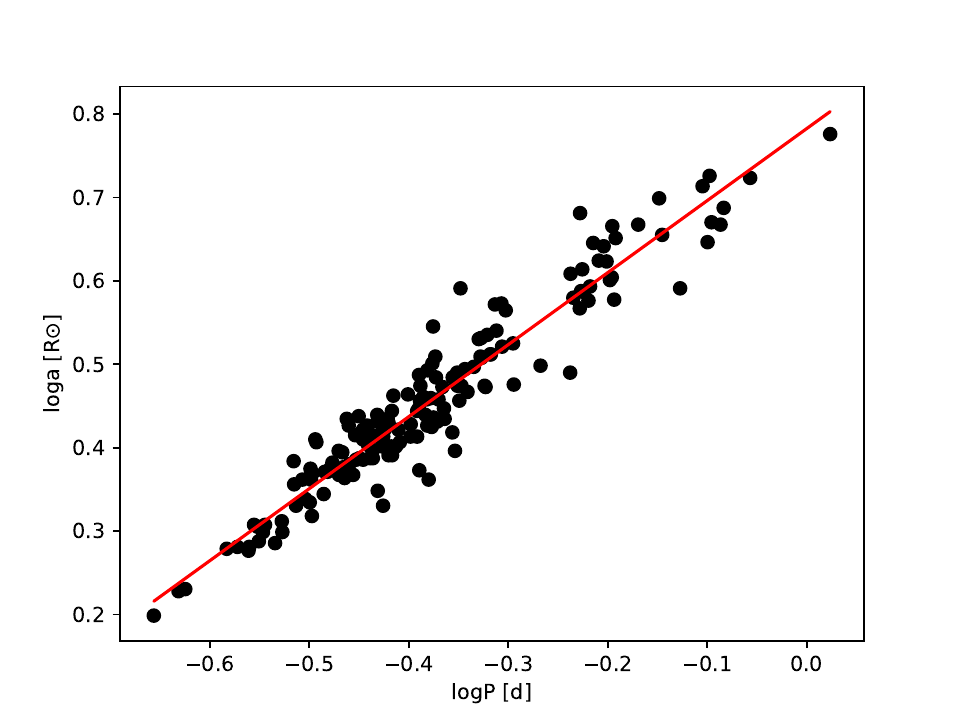}
 \centering
 \caption{The relationship between P and a of the contact binary stars from \citep{2021AJ....162...13L}.}
 \label{p-a}
\end{figure*}

By examining the 168 contact binaries with temperatures below 10000 K from \cite{2021AJ....162...13L}, which were analyzed using both radial velocity curves and photometric light curves, we derived the relationship between the semi-major axis $a$ and the period $P$ \citep{2022NewA...9101695Y}.
The relationship is displayed in Figure \ref{p-a}, the equation is as follows:
\begin{equation}
log a=0.864(\pm0.020) log P + 0.783(\pm0.008).
\end{equation}
When we take the logarithm of Kepler's third law, we can find a constant term related to mass, and 0.783 corresponds to this constant term.
According to this relationship, Phoebe was used to calculate the absolute parameters of these targets. The results are listed in Table \ref{tab:absolute parameters}. 
% \begin{equation}
% M_{T} = 0.0134 \times a^{3} / P^{2},
% \end{equation}

% \begin{equation}
% M_{1} = M_{T} / (1+q) , M_{2} = M_{1} \times q,
% \end{equation}

% \begin{equation}
% R_{1}=r_{1}\times a, R_{2}=r_{2}\times a,
% \end{equation}

% \begin{equation}
% L_{1}=R_{1}^{2} \times (T_{1}/T_{\odot})^{4}L_{\odot}, L_{2}=R_{2}^{2} \times (T_{2}/T_{\odot})^{4}L_{\odot},
% \end{equation}
% where $M_{T}$ is the total mass of the system, and $T_{\odot}$ is the temperature of the Sun. The results are listed in Table \ref{tab:absolute parameters}. 

\begin{figure*}
	\begin{minipage}{0.4999\textwidth}
		\hspace{0.5pt}
        %这个图片路径替换成你的图片路径即可使用
		\centerline{\includegraphics[width=\linewidth]{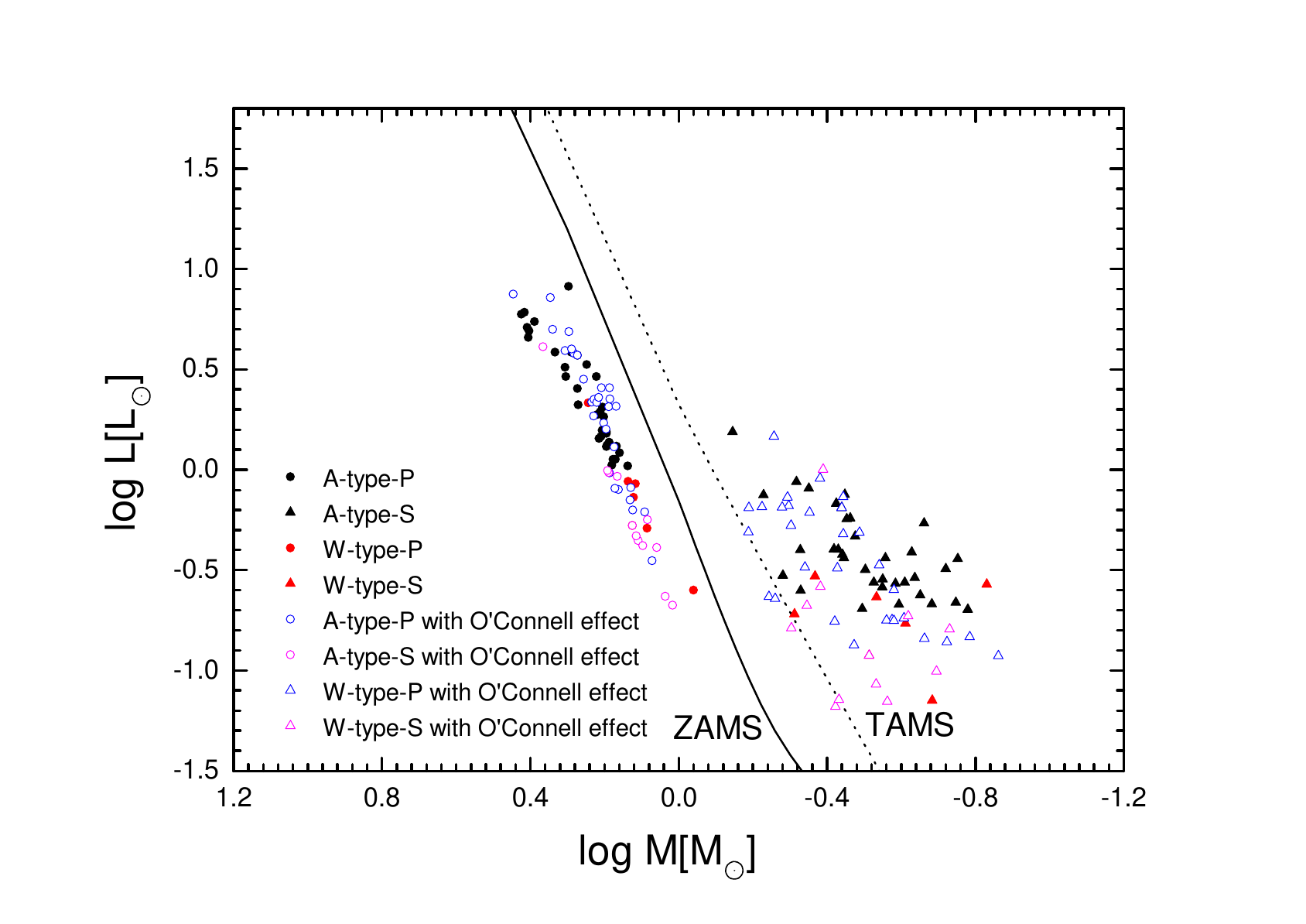}}
          % 加入对这列的图片说明
		%\centerline{Image}
	\end{minipage}
	\begin{minipage}{0.4999\textwidth}
		\hspace{0.5pt}
		\centerline{\includegraphics[width=\linewidth]{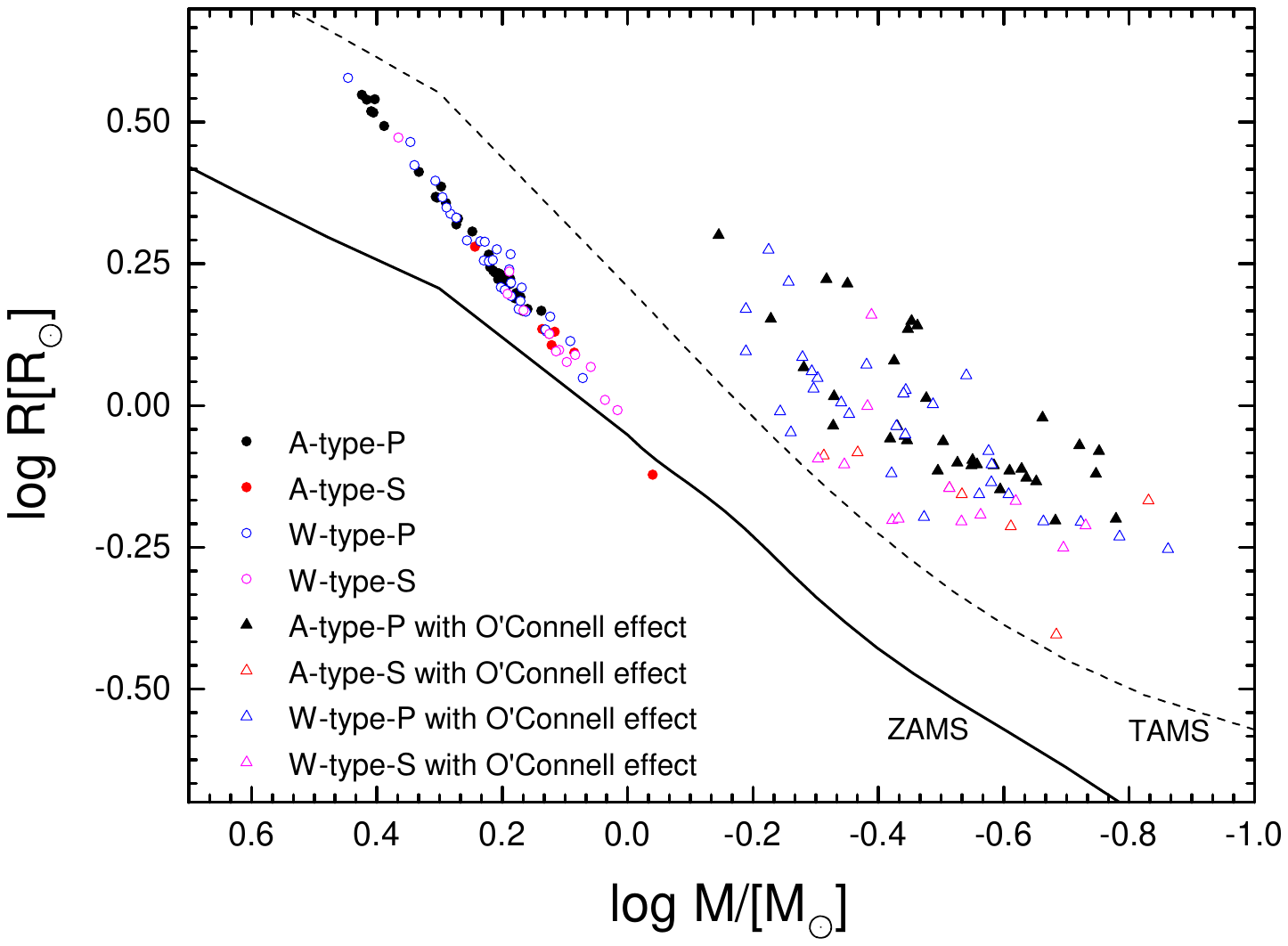}}
		%\centerline{Image}
	\end{minipage}
	\caption{The M-L and M-R relations for 84 contact binaries.}
	\label{M-L-R}
\end{figure*}

In order to study the evolutionary states of  these binary stars, the mass–radius (M-R) and the mass–luminosity (M-L) diagrams were plotted in Figure \ref{M-L-R}. ZAMS line refers to the zero age main sequence line, while TAMS line refers to the terminal age main sequence line, which are constructed by the binary evolutionary code provided by \cite{2002MNRAS.329..897H}. 
Solid dots represent the primary stars, and solid triangles represent the secondary stars. 
The black and red solid dots and triangles represent the primary and secondary stars of A-subtype and W-subtype stars, respectively.
From the M-L diagram, it can be seen that most of the primary stars are close to the ZAMS line, while the secondary stars are located above the TAMS line. The primary reason that most of the primary stars are below the ZAMS line is likely due to systematic temperature errors, which in turn affect the luminosity.
Similarly, in the M-R diagram, the primary stars are between the ZAMS and TAMS lines, and the secondary stars are located above the TAMS line.
Compared to the same mass main sequence stars, the secondary stars have higher luminosities and larger radii, which may be due to energy and mass transfer from the primary components to the secondary ones.
By observing the A-subtype and W-subtype contact binaries in Figure \ref{M-L-R}, we found that for the same mass, the luminosity and radius of each component of the A-subtype contact binaries are greater than those of W-subtype ones, indicating that A-subtype and W-subtype contact systems have different evolutionary paths, and A-subtype stars may be at a later stage of evolution \citep{1996A&A...311..523M, 2008MNRAS.390.1577G}.

% The strength of O’Connell effect is usually represented by $\Delta m$:
% \begin{equation}
% \Delta m=max.\uppercase\expandafter{\romannumeral2} +max. \uppercase\expandafter{\romannumeral1}
% \end{equation}
% $max.\uppercase\expandafter{\romannumeral1}$, $max.\uppercase\expandafter{\romannumeral2}$ corresponds to the maximum after the primary and secondary minimum. The $\lvert \Delta m \rvert $ of these 13 targets is less than 0.04 mag, of which 5 targets are less than 0.001 mag. 

\begin{figure*}
\begin{minipage}{0.53\textwidth}
		\hspace{0.5pt}
        %这个图片路径替换成你的图片路径即可使用
		\centerline{\includegraphics[width=\linewidth]{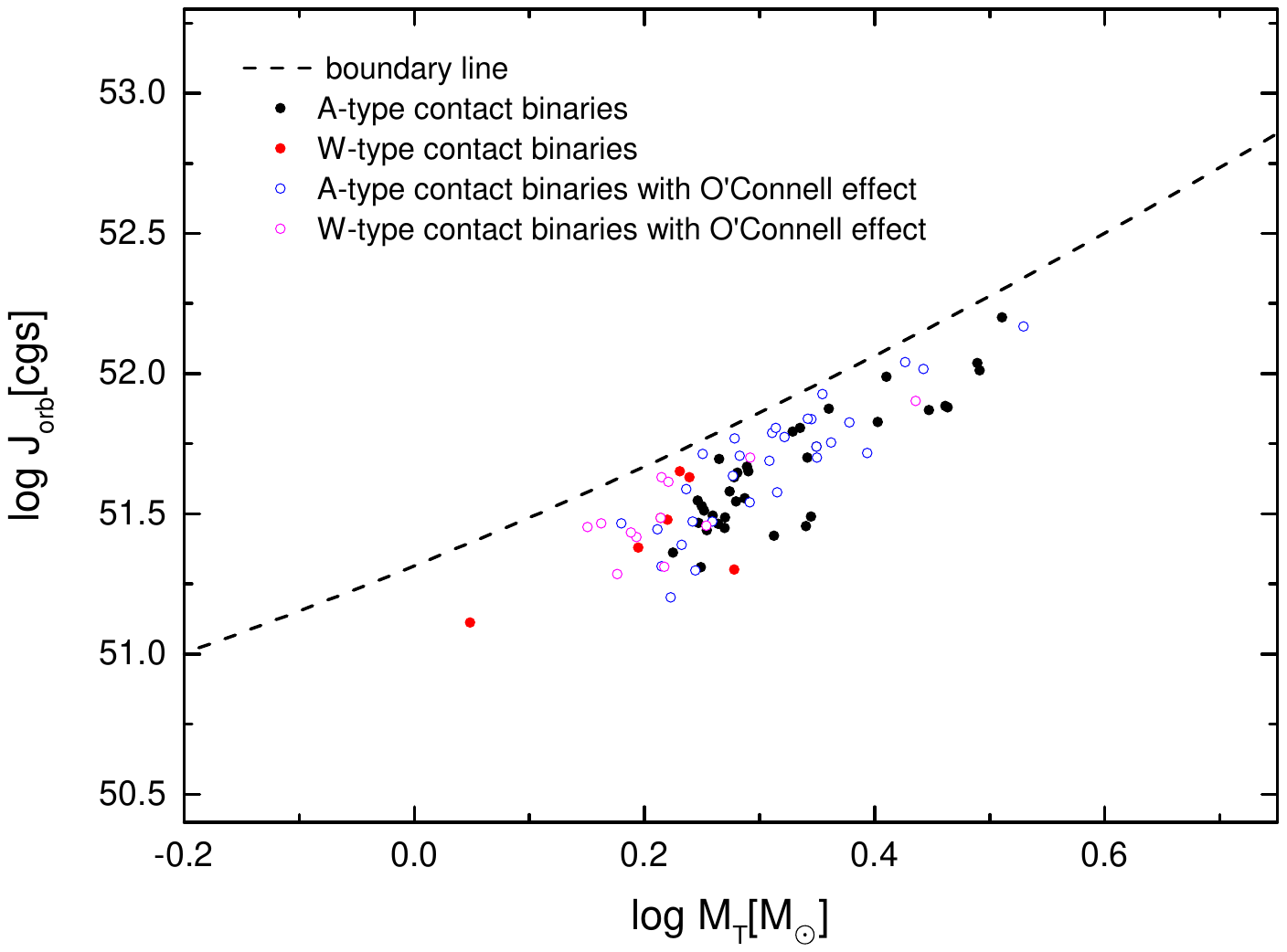}}
          % 加入对这列的图片说明
		%\centerline{Image}
	\end{minipage}
	\begin{minipage}{0.53\linewidth}
		\hspace{0.5pt}
		\centerline{\includegraphics[width=\linewidth]{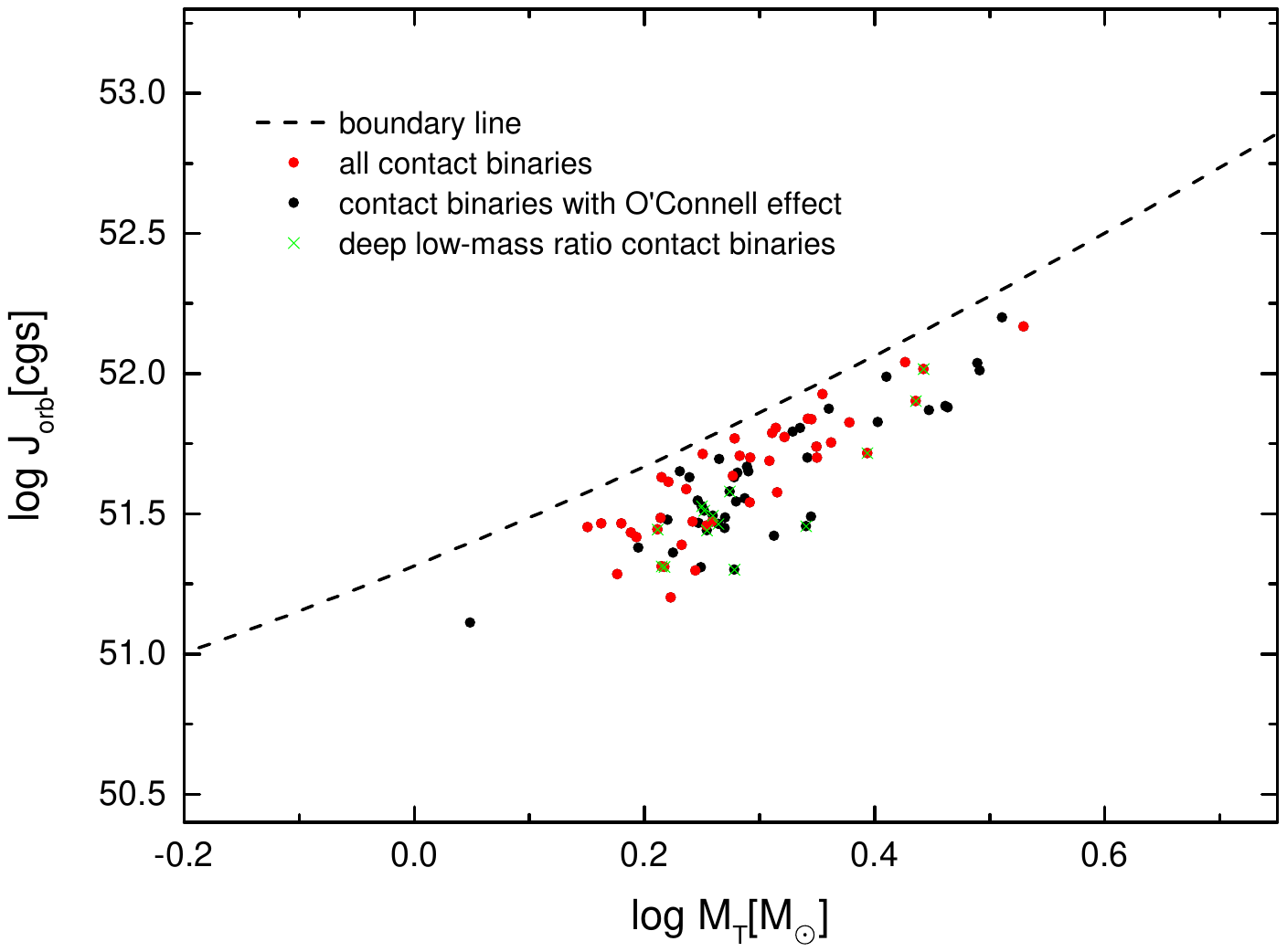}}
		%\centerline{Image}
	\end{minipage}
	\caption{The $M_{T}$ and $J_{orb}$ relations for 84 contact binaries.}
	\label{M-J}
\end{figure*}

The orbital angular momentum was calculated according to the following equation provided by \cite{2006MNRAS.373.1483E}:

\begin{equation}
J_{orb}=1.24 \times 10^{52} \times M_{T}^{5/3} \times P^{1/3} \times q \times (1+q)^{-2}
\end{equation}
The values are displayed in Table \ref{tab:absolute parameters}. The $J_{orb}$-$M_{T}$ diagram is displayed in Figure \ref{M-J}. In the left panel of this figure, the red and black dots represent A-subtype and W-subtype contact binaries, respectively. The boundary line indicates the separation between contact binaries and detached binaries. 
From the left panel of Figure \ref{M-J}, it can be seen that for A-subtype and W-subtype stars with the same mass, most A-subtype instances exhibit lower orbital angular momentum than W-subtype stars. 
This may indicate that the evolutionary pathways of these two types of stars could be different. A-subtype stars are more evolved and are in a later stage of evolution compared to W-subtype stars, therefore they have a smaller orbital angular momentum \citep{1978ApJ...224..885W, 2020MNRAS.492.4112Z, 2021AJ....162...13L}.

Low-mass ratio deep contact binaries are considered to be at the late evolution stages of contact binary and may eventually merge into single stars. In the right panel of Figure \ref{M-J}, the green crosses represent deep, low-mass ratio contact binaries. Compared to other contact binaries, their lower angular momentum suggests that they might be at the late evolution stages of contact binary and are regarded as candidates for binary mergers.

The initial mass is a crucial parameter in determining the evolution and structure of a contact binary system. To further investigate the evolutionary states of these 84 systems, we calculated their initial masses. The initial masses were calculated according to the following equations \citep{2013MNRAS.430.2029Y}:

\begin{equation}
M_{2i}=M_{2}+\Delta M =M_{2}+2.5(M_{L}-M_{2}-0.07)^{0.64}
\end{equation}

\begin{equation}
M_{1i}=M_{1}-(\Delta M-M_{lost})=M_{1}-\Delta M(1-\gamma)
\end{equation}
where $M_{1i}$, $M_{2i}$ are initial masses of primary and secondary star. $M_{1}$, $M_{2}$ are current masses. $M_{L}$ is the mass from the mass-luminosity relation:
$(L_{2}/1.49)^{1/4.216}$. $M_{lost}$ refers to the mass of system loss. $\gamma$ is the ratio of $M_{lost}$ to $\Delta M$, and a value of 0.64 was adopted from \cite{2013MNRAS.430.2029Y}. 
% The ages were estimated according to the following equations from \cite{2014MNRAS.437..185Y}:
% \begin{equation}
% \tau=\tau_{MS}(M_{2i})+\tau_{MS}(\overline{M_{2}})
% \end{equation}

% \begin{equation}
% \tau_{MS}=\frac{10}{(M/M_{\odot})^{4.05}}\times(5.60\times10^{-3}(\frac{M}{M_{\odot}}+3.993)^{3.16}+0.042) \quad Gyr
% \end{equation}
% $\overline{M_{2}}$ refers to the mean value of $M_{2i}$ and $M_{L}$. The calculation results of the initial masses of two components and their ages are shown in Table \ref{tab:absolute parameters}.  

Due to the values of $\delta M=M_{L}-M_{2}$ being less than 0.35 $M_{\odot}$ for some W-subtype stars, and as noted in \cite{2013MNRAS.430.2029Y}, when $\delta M$ for W-subtype stars is less than 0.35 $M_{\odot}$, they have a different value of $\gamma$ and may have different structure in the early evolutionary stage, so the above calculation method is no longer applicable to these targets.
The average initial mass of A-subtype's secondary stars is 1.47 $M_{\odot}$, while that of W-subtype stars is 1.17 $M_{\odot}$.
% \textbf{The average initial mass of A-subtype's secondary stars is 1.99 $M_{\odot}$, while that of W-subtype stars is 1.73 $M_{\odot}$.} This is consistent with the idea that if the initial mass of the secondary is greater than 1.8 $M_{\odot}$, the star will evolve into an A-subtype star; otherwise, it will evolve into a W-subtype star \citep{2013MNRAS.430.2029Y}.
The initial mass of the secondary of A-subtype is greater than that of W-subtype star, this result is consistent with \cite{2013MNRAS.430.2029Y}.
The likely reason is that the secondary stars of the A-subtype have a larger initial mass, which leads to more rapid nuclear reaction processes during their evolution.
This results in a quicker depletion of hydrogen fuel in their cores and a more expedited progression to the later stages of stellar evolution, consequently exhibiting larger radii and higher luminosities, as well as reduced orbital angular momentum. In contrast, stars of the W-subtype likely originated from those with lower initial masses, leading to an extended period of evolution in the main sequence phase, thus they are older \citep{2020MNRAS.492.4112Z, 2021AJ....162...13L}.

We conducted a historical statistical study of these 84 targets and found that 44 of them have been previously studied. Among these, 27 targets were studied by \cite{2023MNRAS.525.4596D}. Therefore, our photometric parameters were compared with those of \cite{2023MNRAS.525.4596D} and other historical studies of these targets. There are 5 parameters for comparison, which are mass ratio $q$, orbital inclination $i$, temperature ratio of primary to secondary stars $T_{2}/T_{1}$, luminosity ratio of primary to secondary stars $L_{2}/L_{1}$, and fillout factor $f$. 
The comparison results are listed in Table \ref{tab:comparison}. 
For the mass ratio q, 21 stars are within 10\% difference. For those targets exceeding the 10\% difference, we conducted a detailed investigation. A total of 23 targets exceeded the 10\% difference, of which 18 targets were from \cite{2023MNRAS.525.4596D}. \cite{2023MNRAS.525.4596D} used machine learning methods to obtain the parameters, which is more efficient for large samples but may result in larger errors for certain individual targets. 
For TIC 267043786 and TIC 207174531, the differences of mass ratios determined from radial velocity observations exceed 10\%. However, we found that the fitting curves of the radial velocity in the historical researches are not good \citep{1986MNRAS.223..581H, 2007AJ....133..169D}, so we re-analyzed the radial velocity curves using PHOEBE with MCMC sampler. The distributions of the mass ratios $q$, semi-major axes $a$, the radial velocity of the system $V_{\gamma}$ and the fitting curve of radial velocity are shown in Figure \ref{267043786} and Figure \ref{207174531} in the appendix. The mass ratio is determined to be 0.331, and the difference is 0.09 for TIC 267043786, while the mass ratio is determined to be 0.441, and the difference is 0.09 for TIC 207174531.
% The historical mass ratio value 0.354 for TIC 267043786 is from \cite{1986MNRAS.223..581H}. We downloaded the radial velocity data of the targets from this paper and recalculated them using PHOEBE with mcmc method. The possible distributions of the mass ratios $q$, semi-major axes $a$, the radial velocity of the system $V_{\gamma}$ and the fitting curve of radial velocity are shown in \ref{267043786} in the appendix. From the figure, the mass ratio is 0.331, and the diff was recalculated as 0.09.
% The historical mass ratio value 0.470 for TIC 207174531 is from \cite{2007AJ....133..169D}. We used the same method as for TIC 267043786, downloaded the data from the paper, and obtained a new mass ratio value of 0.441. The diff was recalculated as 0.09. The possible distribution and the fitting curve of radial velocity are shown in Figure \ref{207174531} in the appendix .
The significant differences between the historical mass ratios and our results for TIC 53842685, 67686223, 261089147 are due to the poor quality of the historical data, which consequently leads to large errors in the mass ratios.
% The historical data of TIC 53842685, 67686223, 261089147 are more scattered compared to the TESS data used in this paper, resulting in larger errors in the derived mass ratios.
For temperature ratio $T_{2}/T_{1}$, 43 of 44 targets are within a 10\% difference range. 
The result of TIC 269505148 is from \cite{2000PASP..112..123R}, the absence of raw data precludes any verification of its validity.
% TIC 269505148 is from \cite{2000PASP..112..123R}, where the paper only provides parameters without data, making it impossible to verify.
For the orbital inclination i, all targets are within 10\% difference. For fillout factor, 17 targets are within 10\% difference. Therefore, the majority of the fillout factors are inconsistent with historical research, which may be related to the mass ratio and data quality.
For luminosity ratio $L_{2}/L_{1}$, 28 targets are within a 10\% difference range. 
The inconsistencies in luminosity ratio for these targets may be related to parameters such as the pass bands, the mass ratio and data quality.

In summary, we conducted a photometric study on 84 contact binaries selected from TESS data and identified 18 deep contact binaries, 39 moderate contact binaries and 27 shallow contact binaries, and among them 43 exhibiting O'Connell effect. The absolute parameters and orbital angular momentum, initial mass, and age were estimated and were used to investigate their evolutionary states. Additionally, We compared our photometric results with other historical results.

% \begin{figure*}
% \centering
% \begin{minipage}[t]{0.49\textwidth}
% \centering
% \includegraphics[width=.99\linewidth]{M-L.pdf}
% \caption{The M-L relations for contact binaries. }
% \label{fig:M-L}
% \end{minipage}
% \begin{minipage}[t]{0.49\textwidth}
% \centering
% \includegraphics[width=.99\linewidth]{M-R.pdf}
% \caption{The M-R relations for contact binaries.}
% \label{fig:M-R}
% \end{minipage}
% \end{figure*}

%% IMPORTANT! The old "\acknowledgment" command has be depreciated. It was
%% not robust enough to handle our new dual anonymous review requirements and
%% thus been replaced with the acknowledgment environment. If you try to 
%% compile with \acknowledgment you will get an error print to the screen
%% and in the compiled pdf.
%% 
%% Also note that the akcnowlodgment environment does not support long amounts of text. If you have a lot of people and institutions to acknowledge, do not use this command. Instead, create a new \section{Acknowledgments}.

\begin{acknowledgments}
We are grateful to the anonymous referees for their constructive comments and suggestions, which significantly improved the quality of this manuscript.

This work was supported by National Natural Science Foundation of China (NSFC) (No. 12273018), and the Joint Research Fund in Astronomy (No.U1931103) under cooperative agreement between NSFC and Chinese Academy of Sciences (CAS), and by the Qilu Young Researcher Project of Shandong University, and by Young Data Scientist Project of the National Astronomical Data Center and by the Cultiation Project for LAMOST Scientific Payoff and Research Achievement of CAMS-CAS, and by the Chinese Academy of Science Interdisciplinary Innovation Team. The calculations in this work were carried out at Supercomputing Center of Shandong University,Weihai. 

The data used in this paper comes from TESS mission. Funding for the TESS mission is provided by NASA Science Mission Directorate. We acknowledge the TESS team for its support of this work.

This work includes data collected by Guoshoujing Telescope (the Large Sky Area Multi-Object Fiber Spectroscopic Telescope LAMOST), which is a national major scientific project built by the Chinese Academy of Sciences. Funding for the project has been provided by the National Development and Reform Commission. 
LAMOST is operated and managed by the National Astronomical Observatories, Chinese Academy of Sciences. 

This work has made use of data from the European Space Agency (ESA) mission Gaia (https://www.cosmos.esa.int/gaia), processed by the Gaia Data Processing and Analysis Consortium (DPAC; https://www.cosmos.esa.int/web/gaia/dpac/consortium). Funding for the DPAC has been provided by national institutions, in particular the institutions participating in the Gaia Multilateral Agreement.
\end{acknowledgments}

\bibliography{sample631}

\begin{thebibliography}{}
\expandafter\ifx\csname natexlab\endcsname\relax\def\natexlab#1{#1}\fi
\providecommand{\url}[1]{\href{#1}{#1}}
\providecommand{\dodoi}[1]{doi:~\href{http://doi.org/#1}{\nolinkurl{#1}}}
\providecommand{\doeprint}[1]{\href{http://ascl.net/#1}{\nolinkurl{http://ascl.net/#1}}}
\providecommand{\doarXiv}[1]{\href{https://arxiv.org/abs/#1}{\nolinkurl{https://arxiv.org/abs/#1}}}

\bibitem[{{Acerbi} {et~al.}(2014){Acerbi}, {Barani}, \& {Martignoni}}]{2014NewA...31....1A}
{Acerbi}, F., {Barani}, C., \& {Martignoni}, M. 2014, \na, 31, 1, \dodoi{10.1016/j.newast.2014.01.013}

\bibitem[{{Arbutina}(2007)}]{2007MNRAS.377.1635A}
{Arbutina}, B. 2007, \mnras, 377, 1635, \dodoi{10.1111/j.1365-2966.2007.11723.x}

\bibitem[{{Arbutina}(2009)}]{2009MNRAS.394..501A}
---. 2009, \mnras, 394, 501, \dodoi{10.1111/j.1365-2966.2008.14332.x}

\bibitem[{{Arbutina} \& {Wadhwa}(2024)}]{2024SerAJ.208....1A}
{Arbutina}, B., \& {Wadhwa}, S. 2024, Serbian Astronomical Journal, 208, 1, \dodoi{10.2298/SAJ2408001A}

\bibitem[{{Binnendijk}(1970)}]{1970VA.....12..217B}
{Binnendijk}, L. 1970, Vistas in Astronomy, 12, 217, \dodoi{10.1016/0083-6656(70)90041-3}

\bibitem[{{Borucki} {et~al.}(2010){Borucki}, {Koch}, {Basri}, {Batalha}, {Brown}, {Caldwell}, {Caldwell}, {Christensen-Dalsgaard}, {Cochran}, {DeVore}, {Dunham}, {Dupree}, {Gautier}, {Geary}, {Gilliland}, {Gould}, {Howell}, {Jenkins}, {Kondo}, {Latham}, {Marcy}, {Meibom}, {Kjeldsen}, {Lissauer}, {Monet}, {Morrison}, {Sasselov}, {Tarter}, {Boss}, {Brownlee}, {Owen}, {Buzasi}, {Charbonneau}, {Doyle}, {Fortney}, {Ford}, {Holman}, {Seager}, {Steffen}, {Welsh}, {Rowe}, {Anderson}, {Buchhave}, {Ciardi}, {Walkowicz}, {Sherry}, {Horch}, {Isaacson}, {Everett}, {Fischer}, {Torres}, {Johnson}, {Endl}, {MacQueen}, {Bryson}, {Dotson}, {Haas}, {Kolodziejczak}, {Van Cleve}, {Chandrasekaran}, {Twicken}, {Quintana}, {Clarke}, {Allen}, {Li}, {Wu}, {Tenenbaum}, {Verner}, {Bruhweiler}, {Barnes}, \& {Prsa}}]{2010Sci...327..977B}
{Borucki}, W.~J., {Koch}, D., {Basri}, G., {et~al.} 2010, Science, 327, 977, \dodoi{10.1126/science.1185402}

\bibitem[{{Bradstreet} \& {Guinan}(1994)}]{1994ASPC...56..228B}
{Bradstreet}, D.~H., \& {Guinan}, E.~F. 1994, in Astronomical Society of the Pacific Conference Series, Vol.~56, Interacting Binary Stars, ed. A.~W. {Shafter}, 228

\bibitem[{{Castelli} \& {Kurucz}(2004)}]{2004A&A...419..725C}
{Castelli}, F., \& {Kurucz}, R.~L. 2004, \aap, 419, 725, \dodoi{10.1051/0004-6361:20040079}

\bibitem[{{Caton} {et~al.}(2019){Caton}, {Gentry}, {Samec}, {Chamberlain}, {Robb}, {Faulkner}, \& {Hill}}]{2019PASP..131e4203C}
{Caton}, D., {Gentry}, D.~R., {Samec}, R.~G., {et~al.} 2019, \pasp, 131, 054203, \dodoi{10.1088/1538-3873/aafb8f}

\bibitem[{{Chen} {et~al.}(2020){Chen}, {Wang}, {Deng}, {de Grijs}, {Yang}, \& {Tian}}]{2020ApJS..249...18C}
{Chen}, X., {Wang}, S., {Deng}, L., {et~al.} 2020, \apjs, 249, 18, \dodoi{10.3847/1538-4365/ab9cae}

\bibitem[{{Chochol} {et~al.}(2001){Chochol}, {van Houten}, {Pribulla}, \& {Grygar}}]{2001CoSka..31....5C}
{Chochol}, D., {van Houten}, C.~J., {Pribulla}, T., \& {Grygar}, J. 2001, Contributions of the Astronomical Observatory Skalnate Pleso, 31, 5

\bibitem[{{Christopoulou} \& {Papageorgiou}(2013)}]{2013AJ....146..157C}
{Christopoulou}, P.~E., \& {Papageorgiou}, A. 2013, \aj, 146, 157, \dodoi{10.1088/0004-6256/146/6/157}

\bibitem[{{Ding} {et~al.}(2023){Ding}, {Ji}, {Li}, {Xiong}, {Cheng}, \& {Wang}}]{2023MNRAS.525.4596D}
{Ding}, X., {Ji}, K., {Li}, X., {et~al.} 2023, \mnras, 525, 4596, \dodoi{10.1093/mnras/stad2565}

\bibitem[{{Duerbeck} \& {Rucinski}(2007)}]{2007AJ....133..169D}
{Duerbeck}, H.~W., \& {Rucinski}, S.~M. 2007, \aj, 133, 169, \dodoi{10.1086/509764}

\bibitem[{{Eggleton} \& {Kiseleva-Eggleton}(2002)}]{2002ApJ...575..461E}
{Eggleton}, P.~P., \& {Kiseleva-Eggleton}, L. 2002, \apj, 575, 461, \dodoi{10.1086/341215}

\bibitem[{{Eker} {et~al.}(2006){Eker}, {Demircan}, {Bilir}, \& {Karata{\c{s}}}}]{2006MNRAS.373.1483E}
{Eker}, Z., {Demircan}, O., {Bilir}, S., \& {Karata{\c{s}}}, Y. 2006, \mnras, 373, 1483, \dodoi{10.1111/j.1365-2966.2006.11073.x}

\bibitem[{{Foreman-Mackey} {et~al.}(2013){Foreman-Mackey}, {Hogg}, {Lang}, \& {Goodman}}]{2013PASP..125..306F}
{Foreman-Mackey}, D., {Hogg}, D.~W., {Lang}, D., \& {Goodman}, J. 2013, \pasp, 125, 306, \dodoi{10.1086/670067}

\bibitem[{{Gaia Collaboration} {et~al.}(2018){Gaia Collaboration}, {Brown}, {Vallenari}, {Prusti}, {de Bruijne}, {Babusiaux}, {Bailer-Jones}, {Biermann}, {Evans}, {Eyer}, {Jansen}, {Jordi}, {Klioner}, {Lammers}, {Lindegren}, {Luri}, {Mignard}, {Panem}, {Pourbaix}, {Randich}, {Sartoretti}, {Siddiqui}, {Soubiran}, {van Leeuwen}, {Walton}, {Arenou}, {Bastian}, {Cropper}, {Drimmel}, {Katz}, {Lattanzi}, {Bakker}, {Cacciari}, {Casta{\~n}eda}, {Chaoul}, {Cheek}, {De Angeli}, {Fabricius}, {Guerra}, {Holl}, {Masana}, {Messineo}, {Mowlavi}, {Nienartowicz}, {Panuzzo}, {Portell}, {Riello}, {Seabroke}, {Tanga}, {Th{\'e}venin}, {Gracia-Abril}, {Comoretto}, {Garcia-Reinaldos}, {Teyssier}, {Altmann}, {Andrae}, {Audard}, {Bellas-Velidis}, {Benson}, {Berthier}, {Blomme}, {Burgess}, {Busso}, {Carry}, {Cellino}, {Clementini}, {Clotet}, {Creevey}, {Davidson}, {De Ridder}, {Delchambre}, {Dell'Oro}, {Ducourant}, {Fern{\'a}ndez-Hern{\'a}ndez}, {Fouesneau}, {Fr{\'e}mat}, {Galluccio}, {Garc{\'\i}a-Torres},
  {Gonz{\'a}lez-N{\'u}{\~n}ez}, {Gonz{\'a}lez-Vidal}, {Gosset}, {Guy}, {Halbwachs}, {Hambly}, {Harrison}, {Hern{\'a}ndez}, {Hestroffer}, {Hodgkin}, {Hutton}, {Jasniewicz}, {Jean-Antoine-Piccolo}, {Jordan}, {Korn}, {Krone-Martins}, {Lanzafame}, {Lebzelter}, {L{\"o}ffler}, {Manteiga}, {Marrese}, {Mart{\'\i}n-Fleitas}, {Moitinho}, {Mora}, {Muinonen}, {Osinde}, {Pancino}, {Pauwels}, {Petit}, {Recio-Blanco}, {Richards}, {Rimoldini}, {Robin}, {Sarro}, {Siopis}, {Smith}, {Sozzetti}, {S{\"u}veges}, {Torra}, {van Reeven}, {Abbas}, {Abreu Aramburu}, {Accart}, {Aerts}, {Altavilla}, {{\'A}lvarez}, {Alvarez}, {Alves}, {Anderson}, {Andrei}, {Anglada Varela}, {Antiche}, {Antoja}, {Arcay}, {Astraatmadja}, {Bach}, {Baker}, {Balaguer-N{\'u}{\~n}ez}, {Balm}, {Barache}, {Barata}, {Barbato}, {Barblan}, {Barklem}, {Barrado}, {Barros}, {Barstow}, {Bartholom{\'e} Mu{\~n}oz}, {Bassilana}, {Becciani}, {Bellazzini}, {Berihuete}, {Bertone}, {Bianchi}, {Bienaym{\'e}}, {Blanco-Cuaresma}, {Boch}, {Boeche}, {Bombrun}, {Borrachero},
  {Bossini}, {Bouquillon}, {Bourda}, {Bragaglia}, {Bramante}, {Breddels}, {Bressan}, {Brouillet}, {Br{\"u}semeister}, {Brugaletta}, {Bucciarelli}, {Burlacu}, {Busonero}, {Butkevich}, {Buzzi}, {Caffau}, {Cancelliere}, {Cannizzaro}, {Cantat-Gaudin}, {Carballo}, {Carlucci}, {Carrasco}, {Casamiquela}, {Castellani}, {Castro-Ginard}, {Charlot}, {Chemin}, {Chiavassa}, {Cocozza}, {Costigan}, {Cowell}, {Crifo}, {Crosta}, {Crowley}, {Cuypers}, {Dafonte}, {Damerdji}, {Dapergolas}, {David}, {David}, {de Laverny}, {De Luise}, {De March}, {de Martino}, {de Souza}, {de Torres}, {Debosscher}, {del Pozo}, {Delbo}, {Delgado}, {Delgado}, {Di Matteo}, {Diakite}, {Diener}, {Distefano}, {Dolding}, {Drazinos}, {Dur{\'a}n}, {Edvardsson}, {Enke}, {Eriksson}, {Esquej}, {Eynard Bontemps}, {Fabre}, {Fabrizio}, {Faigler}, {Falc{\~a}o}, {Farr{\`a}s Casas}, {Federici}, {Fedorets}, {Fernique}, {Figueras}, {Filippi}, {Findeisen}, {Fonti}, {Fraile}, {Fraser}, {Fr{\'e}zouls}, {Gai}, {Galleti}, {Garabato}, {Garc{\'\i}a-Sedano}, {Garofalo},
  {Garralda}, {Gavel}, {Gavras}, {Gerssen}, {Geyer}, {Giacobbe}, {Gilmore}, {Girona}, {Giuffrida}, {Glass}, {Gomes}, {Granvik}, {Gueguen}, {Guerrier}, {Guiraud}, {Guti{\'e}rrez-S{\'a}nchez}, {Haigron}, {Hatzidimitriou}, {Hauser}, {Haywood}, {Heiter}, {Helmi}, {Heu}, {Hilger}, {Hobbs}, {Hofmann}, {Holland}, {Huckle}, {Hypki}, {Icardi}, {Jan{\ss}en}, {Jevardat de Fombelle}, {Jonker}, {Juh{\'a}sz}, {Julbe}, {Karampelas}, {Kewley}, {Klar}, {Kochoska}, {Kohley}, {Kolenberg}, {Kontizas}, {Kontizas}, {Koposov}, {Kordopatis}, {Kostrzewa-Rutkowska}, {Koubsky}, {Lambert}, {Lanza}, {Lasne}, {Lavigne}, {Le Fustec}, {Le Poncin-Lafitte}, {Lebreton}, {Leccia}, {Leclerc}, {Lecoeur-Taibi}, {Lenhardt}, {Leroux}, {Liao}, {Licata}, {Lindstr{\o}m}, {Lister}, {Livanou}, {Lobel}, {L{\'o}pez}, {Managau}, {Mann}, {Mantelet}, {Marchal}, {Marchant}, {Marconi}, {Marinoni}, {Marschalk{\'o}}, {Marshall}, {Martino}, {Marton}, {Mary}, {Massari}, {Matijevi{\v{c}}}, {Mazeh}, {McMillan}, {Messina}, {Michalik}, {Millar}, {Molina}, {Molinaro},
  {Moln{\'a}r}, {Montegriffo}, {Mor}, {Morbidelli}, {Morel}, {Morris}, {Mulone}, {Muraveva}, {Musella}, {Nelemans}, {Nicastro}, {Noval}, {O'Mullane}, {Ord{\'e}novic}, {Ord{\'o}{\~n}ez-Blanco}, {Osborne}, {Pagani}, {Pagano}, {Pailler}, {Palacin}, {Palaversa}, {Panahi}, {Pawlak}, {Piersimoni}, {Pineau}, {Plachy}, {Plum}, {Poggio}, {Poujoulet}, {Pr{\v{s}}a}, {Pulone}, {Racero}, {Ragaini}, {Rambaux}, {Ramos-Lerate}, {Regibo}, {Reyl{\'e}}, {Riclet}, {Ripepi}, {Riva}, {Rivard}, {Rixon}, {Roegiers}, {Roelens}, {Romero-G{\'o}mez}, {Rowell}, {Royer}, {Ruiz-Dern}, {Sadowski}, {Sagrist{\`a} Sell{\'e}s}, {Sahlmann}, {Salgado}, {Salguero}, {Sanna}, {Santana-Ros}, {Sarasso}, {Savietto}, {Schultheis}, {Sciacca}, {Segol}, {Segovia}, {S{\'e}gransan}, {Shih}, {Siltala}, {Silva}, {Smart}, {Smith}, {Solano}, {Solitro}, {Sordo}, {Soria Nieto}, {Souchay}, {Spagna}, {Spoto}, {Stampa}, {Steele}, {Steidelm{\"u}ller}, {Stephenson}, {Stoev}, {Suess}, {Surdej}, {Szabados}, {Szegedi-Elek}, {Tapiador}, {Taris}, {Tauran}, {Taylor},
  {Teixeira}, {Terrett}, {Teyssandier}, {Thuillot}, {Titarenko}, {Torra Clotet}, {Turon}, {Ulla}, {Utrilla}, {Uzzi}, {Vaillant}, {Valentini}, {Valette}, {van Elteren}, {Van Hemelryck}, {van Leeuwen}, {Vaschetto}, {Vecchiato}, {Veljanoski}, {Viala}, {Vicente}, {Vogt}, {von Essen}, {Voss}, {Votruba}, {Voutsinas}, {Walmsley}, {Weiler}, {Wertz}, {Wevers}, {Wyrzykowski}, {Yoldas}, {{\v{Z}}erjal}, {Ziaeepour}, {Zorec}, {Zschocke}, {Zucker}, {Zurbach}, \& {Zwitter}}]{2018A&A...616A...1G}
{Gaia Collaboration}, {Brown}, A.~G.~A., {Vallenari}, A., {et~al.} 2018, \aap, 616, A1, \dodoi{10.1051/0004-6361/201833051}

\bibitem[{{Gaia Collaboration} {et~al.}(2021){Gaia Collaboration}, {Brown}, {Vallenari}, {Prusti}, {de Bruijne}, {Babusiaux}, {Biermann}, {Creevey}, {Evans}, {Eyer}, {Hutton}, {Jansen}, {Jordi}, {Klioner}, {Lammers}, {Lindegren}, {Luri}, {Mignard}, {Panem}, {Pourbaix}, {Randich}, {Sartoretti}, {Soubiran}, {Walton}, {Arenou}, {Bailer-Jones}, {Bastian}, {Cropper}, {Drimmel}, {Katz}, {Lattanzi}, {van Leeuwen}, {Bakker}, {Cacciari}, {Casta{\~n}eda}, {De Angeli}, {Ducourant}, {Fabricius}, {Fouesneau}, {Fr{\'e}mat}, {Guerra}, {Guerrier}, {Guiraud}, {Jean-Antoine Piccolo}, {Masana}, {Messineo}, {Mowlavi}, {Nicolas}, {Nienartowicz}, {Pailler}, {Panuzzo}, {Riclet}, {Roux}, {Seabroke}, {Sordo}, {Tanga}, {Th{\'e}venin}, {Gracia-Abril}, {Portell}, {Teyssier}, {Altmann}, {Andrae}, {Bellas-Velidis}, {Benson}, {Berthier}, {Blomme}, {Brugaletta}, {Burgess}, {Busso}, {Carry}, {Cellino}, {Cheek}, {Clementini}, {Damerdji}, {Davidson}, {Delchambre}, {Dell'Oro}, {Fern{\'a}ndez-Hern{\'a}ndez}, {Galluccio}, {Garc{\'\i}a-Lario},
  {Garcia-Reinaldos}, {Gonz{\'a}lez-N{\'u}{\~n}ez}, {Gosset}, {Haigron}, {Halbwachs}, {Hambly}, {Harrison}, {Hatzidimitriou}, {Heiter}, {Hern{\'a}ndez}, {Hestroffer}, {Hodgkin}, {Holl}, {Jan{\ss}en}, {Jevardat de Fombelle}, {Jordan}, {Krone-Martins}, {Lanzafame}, {L{\"o}ffler}, {Lorca}, {Manteiga}, {Marchal}, {Marrese}, {Moitinho}, {Mora}, {Muinonen}, {Osborne}, {Pancino}, {Pauwels}, {Petit}, {Recio-Blanco}, {Richards}, {Riello}, {Rimoldini}, {Robin}, {Roegiers}, {Rybizki}, {Sarro}, {Siopis}, {Smith}, {Sozzetti}, {Ulla}, {Utrilla}, {van Leeuwen}, {van Reeven}, {Abbas}, {Abreu Aramburu}, {Accart}, {Aerts}, {Aguado}, {Ajaj}, {Altavilla}, {{\'A}lvarez}, {{\'A}lvarez Cid-Fuentes}, {Alves}, {Anderson}, {Anglada Varela}, {Antoja}, {Audard}, {Baines}, {Baker}, {Balaguer-N{\'u}{\~n}ez}, {Balbinot}, {Balog}, {Barache}, {Barbato}, {Barros}, {Barstow}, {Bartolom{\'e}}, {Bassilana}, {Bauchet}, {Baudesson-Stella}, {Becciani}, {Bellazzini}, {Bernet}, {Bertone}, {Bianchi}, {Blanco-Cuaresma}, {Boch}, {Bombrun}, {Bossini},
  {Bouquillon}, {Bragaglia}, {Bramante}, {Breedt}, {Bressan}, {Brouillet}, {Bucciarelli}, {Burlacu}, {Busonero}, {Butkevich}, {Buzzi}, {Caffau}, {Cancelliere}, {C{\'a}novas}, {Cantat-Gaudin}, {Carballo}, {Carlucci}, {Carnerero}, {Carrasco}, {Casamiquela}, {Castellani}, {Castro-Ginard}, {Castro Sampol}, {Chaoul}, {Charlot}, {Chemin}, {Chiavassa}, {Cioni}, {Comoretto}, {Cooper}, {Cornez}, {Cowell}, {Crifo}, {Crosta}, {Crowley}, {Dafonte}, {Dapergolas}, {David}, {David}, {de Laverny}, {De Luise}, {De March}, {De Ridder}, {de Souza}, {de Teodoro}, {de Torres}, {del Peloso}, {del Pozo}, {Delbo}, {Delgado}, {Delgado}, {Delisle}, {Di Matteo}, {Diakite}, {Diener}, {Distefano}, {Dolding}, {Eappachen}, {Edvardsson}, {Enke}, {Esquej}, {Fabre}, {Fabrizio}, {Faigler}, {Fedorets}, {Fernique}, {Fienga}, {Figueras}, {Fouron}, {Fragkoudi}, {Fraile}, {Franke}, {Gai}, {Garabato}, {Garcia-Gutierrez}, {Garc{\'\i}a-Torres}, {Garofalo}, {Gavras}, {Gerlach}, {Geyer}, {Giacobbe}, {Gilmore}, {Girona}, {Giuffrida}, {Gomel}, {Gomez},
  {Gonzalez-Santamaria}, {Gonz{\'a}lez-Vidal}, {Granvik}, {Guti{\'e}rrez-S{\'a}nchez}, {Guy}, {Hauser}, {Haywood}, {Helmi}, {Hidalgo}, {Hilger}, {H{\l}adczuk}, {Hobbs}, {Holland}, {Huckle}, {Jasniewicz}, {Jonker}, {Juaristi Campillo}, {Julbe}, {Karbevska}, {Kervella}, {Khanna}, {Kochoska}, {Kontizas}, {Kordopatis}, {Korn}, {Kostrzewa-Rutkowska}, {Kruszy{\'n}ska}, {Lambert}, {Lanza}, {Lasne}, {Le Campion}, {Le Fustec}, {Lebreton}, {Lebzelter}, {Leccia}, {Leclerc}, {Lecoeur-Taibi}, {Liao}, {Licata}, {Lindstr{\o}m}, {Lister}, {Livanou}, {Lobel}, {Madrero Pardo}, {Managau}, {Mann}, {Marchant}, {Marconi}, {Marcos Santos}, {Marinoni}, {Marocco}, {Marshall}, {Martin Polo}, {Mart{\'\i}n-Fleitas}, {Masip}, {Massari}, {Mastrobuono-Battisti}, {Mazeh}, {McMillan}, {Messina}, {Michalik}, {Millar}, {Mints}, {Molina}, {Molinaro}, {Moln{\'a}r}, {Montegriffo}, {Mor}, {Morbidelli}, {Morel}, {Morris}, {Mulone}, {Munoz}, {Muraveva}, {Murphy}, {Musella}, {Noval}, {Ord{\'e}novic}, {Orr{\`u}}, {Osinde}, {Pagani}, {Pagano},
  {Palaversa}, {Palicio}, {Panahi}, {Pawlak}, {Pe{\~n}alosa Esteller}, {Penttil{\"a}}, {Piersimoni}, {Pineau}, {Plachy}, {Plum}, {Poggio}, {Poretti}, {Poujoulet}, {Pr{\v{s}}a}, {Pulone}, {Racero}, {Ragaini}, {Rainer}, {Raiteri}, {Rambaux}, {Ramos}, {Ramos-Lerate}, {Re Fiorentin}, {Regibo}, {Reyl{\'e}}, {Ripepi}, {Riva}, {Rixon}, {Robichon}, {Robin}, {Roelens}, {Rohrbasser}, {Romero-G{\'o}mez}, {Rowell}, {Royer}, {Rybicki}, {Sadowski}, {Sagrist{\`a} Sell{\'e}s}, {Sahlmann}, {Salgado}, {Salguero}, {Samaras}, {Sanchez Gimenez}, {Sanna}, {Santove{\~n}a}, {Sarasso}, {Schultheis}, {Sciacca}, {Segol}, {Segovia}, {S{\'e}gransan}, {Semeux}, {Shahaf}, {Siddiqui}, {Siebert}, {Siltala}, {Slezak}, {Smart}, {Solano}, {Solitro}, {Souami}, {Souchay}, {Spagna}, {Spoto}, {Steele}, {Steidelm{\"u}ller}, {Stephenson}, {S{\"u}veges}, {Szabados}, {Szegedi-Elek}, {Taris}, {Tauran}, {Taylor}, {Teixeira}, {Thuillot}, {Tonello}, {Torra}, {Torra}, {Turon}, {Unger}, {Vaillant}, {van Dillen}, {Vanel}, {Vecchiato}, {Viala}, {Vicente},
  {Voutsinas}, {Weiler}, {Wevers}, {Wyrzykowski}, {Yoldas}, {Yvard}, {Zhao}, {Zorec}, {Zucker}, {Zurbach}, \& {Zwitter}}]{2021A&A...649A...1G}
---. 2021, \aap, 649, A1, \dodoi{10.1051/0004-6361/202039657}

\bibitem[{{Gazeas} \& {St{\c{e}}pie{\'n}}(2008)}]{2008MNRAS.390.1577G}
{Gazeas}, K., \& {St{\c{e}}pie{\'n}}, K. 2008, \mnras, 390, 1577, \dodoi{10.1111/j.1365-2966.2008.13844.x}

\bibitem[{{Gezer} \& {Bozkurt}(2016)}]{2016NewA...44...40G}
{Gezer}, {\.I}., \& {Bozkurt}, Z. 2016, \na, 44, 40, \dodoi{10.1016/j.newast.2015.09.009}

\bibitem[{{Guinan} \& {Bradstreet}(1988)}]{1988ASIC..241..345G}
{Guinan}, E.~F., \& {Bradstreet}, D.~H. 1988, in NATO Advanced Study Institute (ASI) Series C, Vol. 241, Formation and Evolution of Low Mass Stars, ed. A.~K. {Dupree} \& M.~T.~V.~T. {Lago}, 345, \dodoi{10.1007/978-94-009-3037-7_23}

\bibitem[{{Hilditch} \& {King}(1986)}]{1986MNRAS.223..581H}
{Hilditch}, R.~W., \& {King}, D.~J. 1986, \mnras, 223, 581, \dodoi{10.1093/mnras/223.3.581}

\bibitem[{{Holland}(1992)}]{1992SciAm.267a..66H}
{Holland}, J.~H. 1992, Scientific American, 267, 66, \dodoi{10.1038/scientificamerican0792-66}

\bibitem[{{Hurley} {et~al.}(2002){Hurley}, {Tout}, \& {Pols}}]{2002MNRAS.329..897H}
{Hurley}, J.~R., {Tout}, C.~A., \& {Pols}, O.~R. 2002, \mnras, 329, 897, \dodoi{10.1046/j.1365-8711.2002.05038.x}

\bibitem[{{Hut}(1980)}]{1980A&A....92..167H}
{Hut}, P. 1980, \aap, 92, 167

\bibitem[{{Jiang} {et~al.}(2010){Jiang}, {Han}, {Wang}, {Jiang}, \& {Li}}]{2010MNRAS.405.2485J}
{Jiang}, D., {Han}, Z., {Wang}, J., {Jiang}, T., \& {Li}, L. 2010, \mnras, 405, 2485, \dodoi{10.1111/j.1365-2966.2010.16615.x}

\bibitem[{{Koch} {et~al.}(2010){Koch}, {Borucki}, {Basri}, {Batalha}, {Brown}, {Caldwell}, {Christensen-Dalsgaard}, {Cochran}, {DeVore}, {Dunham}, {Gautier}, {Geary}, {Gilliland}, {Gould}, {Jenkins}, {Kondo}, {Latham}, {Lissauer}, {Marcy}, {Monet}, {Sasselov}, {Boss}, {Brownlee}, {Caldwell}, {Dupree}, {Howell}, {Kjeldsen}, {Meibom}, {Morrison}, {Owen}, {Reitsema}, {Tarter}, {Bryson}, {Dotson}, {Gazis}, {Haas}, {Kolodziejczak}, {Rowe}, {Van Cleve}, {Allen}, {Chandrasekaran}, {Clarke}, {Li}, {Quintana}, {Tenenbaum}, {Twicken}, \& {Wu}}]{2010ApJ...713L..79K}
{Koch}, D.~G., {Borucki}, W.~J., {Basri}, G., {et~al.} 2010, \apjl, 713, L79, \dodoi{10.1088/2041-8205/713/2/L79}

\bibitem[{{Lapasset}(1980)}]{1980AJ.....85.1098L}
{Lapasset}, E. 1980, \aj, 85, 1098, \dodoi{10.1086/112774}

\bibitem[{{Lee} {et~al.}(2013){Lee}, {Hinse}, \& {Park}}]{2013AJ....145..100L}
{Lee}, J.~W., {Hinse}, T.~C., \& {Park}, J.-H. 2013, \aj, 145, 100, \dodoi{10.1088/0004-6256/145/4/100}

\bibitem[{{Lee} {et~al.}(2009){Lee}, {Kim}, {Lee}, \& {Youn}}]{2009PASP..121.1366L}
{Lee}, J.~W., {Kim}, S.-L., {Lee}, C.-U., \& {Youn}, J.-H. 2009, \pasp, 121, 1366, \dodoi{10.1086/649230}

\bibitem[{{Li} {et~al.}(2016){Li}, {Gao}, {Hu}, {Guo}, {Jiang}, \& {Chen}}]{2016Ap&SS.361...63L}
{Li}, K., {Gao}, D.~Y., {Hu}, S.~M., {et~al.} 2016, \apss, 361, 63, \dodoi{10.1007/s10509-016-2649-8}

\bibitem[{{Li} {et~al.}(2022){Li}, {Gao}, {Liu}, {Gao}, {Li}, {Chen}, \& {Sun}}]{2022AJ....164..202L}
{Li}, K., {Gao}, X., {Liu}, X.-Y., {et~al.} 2022, \aj, 164, 202, \dodoi{10.3847/1538-3881/ac8ff2}

\bibitem[{{Li} {et~al.}(2015){Li}, {Hu}, {Guo}, {Jiang}, {Gao}, \& {Chen}}]{2015NewA...41...17L}
{Li}, K., {Hu}, S., {Guo}, D., {et~al.} 2015, \na, 41, 17, \dodoi{10.1016/j.newast.2015.04.010}

\bibitem[{{Li} {et~al.}(2014){Li}, {Hu}, {Jiang}, {Chen}, \& {Ren}}]{2014NewA...30...64L}
{Li}, K., {Hu}, S.~M., {Jiang}, Y.~G., {Chen}, X., \& {Ren}, D.~Y. 2014, \na, 30, 64, \dodoi{10.1016/j.newast.2014.01.004}

\bibitem[{{Li} \& {Qian}(2013)}]{2013NewA...22...57L}
{Li}, K., \& {Qian}, S.~B. 2013, \na, 22, 57, \dodoi{10.1016/j.newast.2013.01.006}

\bibitem[{{Li} {et~al.}(2021{\natexlab{a}}){Li}, {Xia}, {Kim}, {Hu}, {Guo}, {Jeong}, {Chen}, \& {Gao}}]{2021ApJ...922..122L}
{Li}, K., {Xia}, Q.-Q., {Kim}, C.-H., {et~al.} 2021{\natexlab{a}}, \apj, 922, 122, \dodoi{10.3847/1538-4357/ac242f}

\bibitem[{{Li} {et~al.}(2019){Li}, {Xia}, {Michel}, {Hu}, {Guo}, {Gao}, {Chen}, \& {Gao}}]{2019MNRAS.485.4588L}
{Li}, K., {Xia}, Q.-Q., {Michel}, R., {et~al.} 2019, \mnras, 485, 4588, \dodoi{10.1093/mnras/stz715}

\bibitem[{{Li} {et~al.}(2021{\natexlab{b}}){Li}, {Xia}, {Kim}, {Gao}, {Hu}, {Guo}, {Gao}, {Chen}, \& {Guo}}]{2021AJ....162...13L}
{Li}, K., {Xia}, Q.-Q., {Kim}, C.-H., {et~al.} 2021{\natexlab{b}}, \aj, 162, 13, \dodoi{10.3847/1538-3881/abfc53}

\bibitem[{{Li} \& {Zhang}(2006)}]{2006MNRAS.369.2001L}
{Li}, L., \& {Zhang}, F. 2006, \mnras, 369, 2001, \dodoi{10.1111/j.1365-2966.2006.10462.x}

\bibitem[{{Lightkurve Collaboration} {et~al.}(2018){Lightkurve Collaboration}, {Cardoso}, {Hedges}, {Gully-Santiago}, {Saunders}, {Cody}, {Barclay}, {Hall}, {Sagear}, {Turtelboom}, {Zhang}, {Tzanidakis}, {Mighell}, {Coughlin}, {Bell}, {Berta-Thompson}, {Williams}, {Dotson}, \& {Barentsen}}]{2018ascl.soft12013L}
{Lightkurve Collaboration}, {Cardoso}, J. V. d.~M., {Hedges}, C., {et~al.} 2018, {Lightkurve: Kepler and TESS time series analysis in Python}, Astrophysics Source Code Library, record ascl:1812.013.
\newblock \doeprint{1812.013}

\bibitem[{{Liu} {et~al.}(1996){Liu}, {Soonthornthum}, {Yang}, {Gu}, {Niparugs}, {Aniwat Sooksawat}, {Wang}, \& {Naksata}}]{1996A&AS..118..453L}
{Liu}, Q., {Soonthornthum}, B., {Yang}, Y., {et~al.} 1996, \aaps, 118, 453

\bibitem[{{Liu} \& {Yang}(2003)}]{2003ChJAA...3..142L}
{Liu}, Q.-Y., \& {Yang}, Y.-L. 2003, \cjaa, 3, 142, \dodoi{10.1088/1009-9271/3/2/142}

\bibitem[{{Liu} {et~al.}(2023){Liu}, {Li}, {Michel}, {Gao}, {Gao}, {Liu}, {Yin}, {Wang}, \& {Sun}}]{2023MNRAS.519.5760L}
{Liu}, X.-Y., {Li}, K., {Michel}, R., {et~al.} 2023, \mnras, 519, 5760, \dodoi{10.1093/mnras/stad026}

\bibitem[{{Lucy}(1967)}]{1967ZA.....65...89L}
{Lucy}, L.~B. 1967, \zap, 65, 89

\bibitem[{{Luo} {et~al.}(2015){Luo}, {Zhao}, {Zhao}, {Deng}, {Liu}, {Jing}, {Wang}, {Zhang}, {Shi}, {Cui}, {Chu}, {Li}, {Bai}, {Wu}, {Cai}, {Cao}, {Cao}, {Carlin}, {Chen}, {Chen}, {Chen}, {Chen}, {Chen}, {Chen}, {Chen}, {Christlieb}, {Chu}, {Cui}, {Dong}, {Du}, {Fan}, {Feng}, {Fu}, {Gao}, {Gong}, {Gu}, {Guo}, {Han}, {He}, {Hou}, {Hou}, {Hou}, {Hu}, {Hu}, {Hu}, {Huo}, {Jia}, {Jiang}, {Jiang}, {Jiang}, {Jin}, {Kong}, {Kong}, {Lei}, {Li}, {Li}, {Li}, {Li}, {Li}, {Li}, {Li}, {Li}, {Li}, {Li}, {Li}, {Li}, {Liang}, {Lin}, {Liu}, {Liu}, {Liu}, {Liu}, {Lu}, {Luo}, {Mao}, {Newberg}, {Ni}, {Qi}, {Qi}, {Shen}, {Shi}, {Song}, {Song}, {Su}, {Su}, {Tang}, {Tao}, {Tian}, {Wang}, {Wang}, {Wang}, {Wang}, {Wang}, {Wang}, {Wang}, {Wang}, {Wang}, {Wang}, {Wang}, {Wang}, {Wang}, {Wang}, {Wang}, {Wang}, {Wang}, {Wang}, {Wang}, {Wang}, {Wei}, {Wei}, {Wu}, {Wu}, {Wu}, {Wu}, {Xing}, {Xu}, {Xu}, {Xu}, {Yan}, {Yang}, {Yang}, {Yang}, {Yang}, {Yao}, {Yu}, {Yuan}, {Yuan}, {Yuan}, {Yuan}, {Zhai}, {Zhang}, {Zhang}, {Zhang}, {Zhang},
  {Zhang}, {Zhang}, {Zhang}, {Zhang}, {Zhao}, {Zhou}, {Zhou}, {Zhu}, {Zhu}, {Zou}, \& {Zuo}}]{2015RAA....15.1095L}
{Luo}, A.~L., {Zhao}, Y.-H., {Zhao}, G., {et~al.} 2015, Research in Astronomy and Astrophysics, 15, 1095, \dodoi{10.1088/1674-4527/15/8/002}

\bibitem[{{Maceroni} {et~al.}(1985){Maceroni}, {Milano}, \& {Russo}}]{1985MNRAS.217..843M}
{Maceroni}, C., {Milano}, L., \& {Russo}, G. 1985, \mnras, 217, 843, \dodoi{10.1093/mnras/217.4.843}

\bibitem[{{Maceroni} \& {van't Veer}(1996)}]{1996A&A...311..523M}
{Maceroni}, C., \& {van't Veer}, F. 1996, \aap, 311, 523

\bibitem[{{Marsh} {et~al.}(2017){Marsh}, {Prince}, {Mahabal}, {Bellm}, {Drake}, \& {Djorgovski}}]{2017MNRAS.465.4678M}
{Marsh}, F.~M., {Prince}, T.~A., {Mahabal}, A.~A., {et~al.} 2017, \mnras, 465, 4678, \dodoi{10.1093/mnras/stw2110}

\bibitem[{{Matijevi{\v{c}}} {et~al.}(2012){Matijevi{\v{c}}}, {Pr{\v{s}}a}, {Orosz}, {Welsh}, {Bloemen}, \& {Barclay}}]{2012AJ....143..123M}
{Matijevi{\v{c}}}, G., {Pr{\v{s}}a}, A., {Orosz}, J.~A., {et~al.} 2012, \aj, 143, 123, \dodoi{10.1088/0004-6256/143/5/123}

\bibitem[{{Nelson} {et~al.}(2002){Nelson}, {Robb}, {Kaiser}, \& {Billings}}]{2002IBVS.5285....1N}
{Nelson}, R.~H., {Robb}, R.~M., {Kaiser}, D.~H., \& {Billings}, G.~B. 2002, Information Bulletin on Variable Stars, 5285, 1

\bibitem[{{O'Connell}(1951)}]{1951PRCO....2...85O}
{O'Connell}, D.~J.~K. 1951, Publications of the Riverview College Observatory, 2, 85

\bibitem[{{Paki} \& {Poro}(2024)}]{2024arXiv240518618P}
{Paki}, E., \& {Poro}, A. 2024, arXiv e-prints, arXiv:2405.18618, \dodoi{10.48550/arXiv.2405.18618}

\bibitem[{{Poro} {et~al.}(2024){Poro}, {Tanriver}, {Michel}, \& {Paki}}]{2024PASP..136b4201P}
{Poro}, A., {Tanriver}, M., {Michel}, R., \& {Paki}, E. 2024, \pasp, 136, 024201, \dodoi{10.1088/1538-3873/ad1ed3}

\bibitem[{{Pr{\v{s}}a} {et~al.}(2016){Pr{\v{s}}a}, {Conroy}, {Horvat}, {Pablo}, {Kochoska}, {Bloemen}, {Giammarco}, {Hambleton}, \& {Degroote}}]{2016ApJS..227...29P}
{Pr{\v{s}}a}, A., {Conroy}, K.~E., {Horvat}, M., {et~al.} 2016, \apjs, 227, 29, \dodoi{10.3847/1538-4365/227/2/29}

\bibitem[{{Pr{\v{s}}a} {et~al.}(2022){Pr{\v{s}}a}, {Kochoska}, {Conroy}, {Eisner}, {Hey}, {IJspeert}, {Kruse}, {Fleming}, {Johnston}, {Kristiansen}, {LaCourse}, {Mortensen}, {Pepper}, {Stassun}, {Torres}, {Abdul-Masih}, {Chakraborty}, {Gagliano}, {Guo}, {Hambleton}, {Hong}, {Jacobs}, {Jones}, {Kostov}, {Lee}, {Omohundro}, {Orosz}, {Page}, {Powell}, {Rappaport}, {Reed}, {Schnittman}, {Schwengeler}, {Shporer}, {Terentev}, {Vanderburg}, {Welsh}, {Caldwell}, {Doty}, {Jenkins}, {Latham}, {Ricker}, {Seager}, {Schlieder}, {Shiao}, {Vanderspek}, \& {Winn}}]{2022ApJS..258...16P}
{Pr{\v{s}}a}, A., {Kochoska}, A., {Conroy}, K.~E., {et~al.} 2022, \apjs, 258, 16, \dodoi{10.3847/1538-4365/ac324a}

\bibitem[{{Pych} {et~al.}(2004){Pych}, {Rucinski}, {DeBond}, {Thomson}, {Capobianco}, {Blake}, {Og{\l}oza}, {Stachowski}, {Rogoziecki}, {Ligeza}, \& {Gazeas}}]{2004AJ....127.1712P}
{Pych}, W., {Rucinski}, S.~M., {DeBond}, H., {et~al.} 2004, \aj, 127, 1712, \dodoi{10.1086/382105}

\bibitem[{{Qian}(2003)}]{2003MNRAS.342.1260Q}
{Qian}, S. 2003, \mnras, 342, 1260, \dodoi{10.1046/j.1365-8711.2003.06627.x}

\bibitem[{{Qian} {et~al.}(2006){Qian}, {Yang}, {Zhu}, {He}, \& {Yuan}}]{2006Ap&SS.304...25Q}
{Qian}, S., {Yang}, Y., {Zhu}, L., {He}, J., \& {Yuan}, J. 2006, \apss, 304, 25, \dodoi{10.1007/s10509-006-9114-z}

\bibitem[{{Qian} {et~al.}(2017){Qian}, {He}, {Zhang}, {Zhu}, {Shi}, {Zhao}, \& {Zhou}}]{2017RAA....17...87Q}
{Qian}, S.-B., {He}, J.-J., {Zhang}, J., {et~al.} 2017, Research in Astronomy and Astrophysics, 17, 087, \dodoi{10.1088/1674-4527/17/8/87}

\bibitem[{{Qian} {et~al.}(2007){Qian}, {Yuan}, {Soonthornthum}, {Zhu}, {He}, \& {Yang}}]{2007ApJ...671..811Q}
{Qian}, S.~B., {Yuan}, J.~Z., {Soonthornthum}, B., {et~al.} 2007, \apj, 671, 811, \dodoi{10.1086/522421}

\bibitem[{{Qian} {et~al.}(2005){Qian}, {Zhu}, {Soonthornthum}, {Yuan}, {Yang}, \& {He}}]{2005AJ....130.1206Q}
{Qian}, S.~B., {Zhu}, L.~Y., {Soonthornthum}, B., {et~al.} 2005, \aj, 130, 1206, \dodoi{10.1086/432544}

\bibitem[{{Qian} {et~al.}(2014){Qian}, {Wang}, {Zhu}, {Snoonthornthum}, {Wang}, {Zhao}, {Zhou}, {Liao}, \& {Liu}}]{2014ApJS..212....4Q}
{Qian}, S.~B., {Wang}, J.~J., {Zhu}, L.~Y., {et~al.} 2014, \apjs, 212, 4, \dodoi{10.1088/0067-0049/212/1/4}

\bibitem[{{Rahman}(2000)}]{2000PASP..112..123R}
{Rahman}, A. 2000, \pasp, 112, 123, \dodoi{10.1086/316485}

\bibitem[{{Rasio}(1995)}]{1995ApJ...444L..41R}
{Rasio}, F.~A. 1995, \apjl, 444, L41, \dodoi{10.1086/187855}

\bibitem[{{Rasio} \& {Shapiro}(1995)}]{1995ApJ...438..887R}
{Rasio}, F.~A., \& {Shapiro}, S.~L. 1995, \apj, 438, 887, \dodoi{10.1086/175130}

\bibitem[{{Ricker} {et~al.}(2015){Ricker}, {Winn}, {Vanderspek}, {Latham}, {Bakos}, {Bean}, {Berta-Thompson}, {Brown}, {Buchhave}, {Butler}, {Butler}, {Chaplin}, {Charbonneau}, {Christensen-Dalsgaard}, {Clampin}, {Deming}, {Doty}, {De Lee}, {Dressing}, {Dunham}, {Endl}, {Fressin}, {Ge}, {Henning}, {Holman}, {Howard}, {Ida}, {Jenkins}, {Jernigan}, {Johnson}, {Kaltenegger}, {Kawai}, {Kjeldsen}, {Laughlin}, {Levine}, {Lin}, {Lissauer}, {MacQueen}, {Marcy}, {McCullough}, {Morton}, {Narita}, {Paegert}, {Palle}, {Pepe}, {Pepper}, {Quirrenbach}, {Rinehart}, {Sasselov}, {Sato}, {Seager}, {Sozzetti}, {Stassun}, {Sullivan}, {Szentgyorgyi}, {Torres}, {Udry}, \& {Villasenor}}]{2015JATIS...1a4003R}
{Ricker}, G.~R., {Winn}, J.~N., {Vanderspek}, R., {et~al.} 2015, Journal of Astronomical Telescopes, Instruments, and Systems, 1, 014003, \dodoi{10.1117/1.JATIS.1.1.014003}

\bibitem[{{Ruci{\'n}ski}(1969)}]{1969AcA....19..245R}
{Ruci{\'n}ski}, S.~M. 1969, \actaa, 19, 245

\bibitem[{{Samec} {et~al.}(2017){Samec}, {Norris}, {Hill}, {van Hamme}, \& {Faulkner}}]{2017JAVSO..45....3S}
{Samec}, R.~G., {Norris}, C.~L., {Hill}, B.~L., {van Hamme}, W., \& {Faulkner}, D.~R. 2017, \jaavso, 45, 3

\bibitem[{{Samec} {et~al.}(2016){Samec}, {Norris}, {Van Hamme}, {Faulkner}, \& {Hill}}]{2016AJ....152..219S}
{Samec}, R.~G., {Norris}, C.~L., {Van Hamme}, W., {Faulkner}, D.~R., \& {Hill}, R.~L. 2016, \aj, 152, 219, \dodoi{10.3847/0004-6256/152/6/219}

\bibitem[{{Shaw} {et~al.}(1990){Shaw}, {Guinan}, \& {Garasi}}]{1990BAAS...22.1296S}
{Shaw}, J.~S., {Guinan}, E.~F., \& {Garasi}, C.~J. 1990, in Bulletin of the American Astronomical Society, Vol.~22, 1296

\bibitem[{{Stepien}(2006)}]{2006AcA....56..347S}
{Stepien}, K. 2006, \actaa, 56, 347, \dodoi{10.48550/arXiv.astro-ph/0701529}

\bibitem[{{Tang} {et~al.}(2022){Tang}, {Guo}, {Li}, {Gai}, \& {Li}}]{2022RAA....22c5009T}
{Tang}, Y.-K., {Guo}, Y.-N., {Li}, K., {Gai}, N., \& {Li}, Z.-K. 2022, Research in Astronomy and Astrophysics, 22, 035009, \dodoi{10.1088/1674-4527/ac4705}

\bibitem[{{Ula{\c{s}}} \& {Ulusoy}(2015)}]{2015NewA...41....1U}
{Ula{\c{s}}}, B., \& {Ulusoy}, C. 2015, \na, 41, 1, \dodoi{10.1016/j.newast.2015.05.001}

\bibitem[{{von Zeipel}(1924)}]{1924MNRAS..84..665V}
{von Zeipel}, H. 1924, \mnras, 84, 665, \dodoi{10.1093/mnras/84.9.665}

\bibitem[{{Wadhwa}(2006)}]{2006Ap&SS.301..195W}
{Wadhwa}, S.~S. 2006, \apss, 301, 195, \dodoi{10.1007/s10509-006-2062-9}

\bibitem[{{Wadhwa} {et~al.}(2021){Wadhwa}, {De Horta}, {Filipovi{\'c}}, {Tothill}, {Arbutina}, {Petrovi{\'c}}, \& {Djura{\v{s}}evi{\'c}}}]{2021MNRAS.501..229W}
{Wadhwa}, S.~S., {De Horta}, A., {Filipovi{\'c}}, M.~D., {et~al.} 2021, \mnras, 501, 229, \dodoi{10.1093/mnras/staa3637}

\bibitem[{{Wang}(1994)}]{1994ApJ...434..277W}
{Wang}, J.~M. 1994, \apj, 434, 277, \dodoi{10.1086/174725}

\bibitem[{{Wang}(1995)}]{1995AJ....110..782W}
{Wang}, J.-M. 1995, \aj, 110, 782, \dodoi{10.1086/117563}

\bibitem[{{Wang} {et~al.}(2024){Wang}, {Li}, {Guo}, {Wang}, {Gao}, {Gao}, \& {Sun}}]{2024ApJ...976..223W}
{Wang}, L.-H., {Li}, K., {Guo}, Y.-N., {et~al.} 2024, \apj, 976, 223, \dodoi{10.3847/1538-4357/ad7f4b}

\bibitem[{{Wilson}(1978)}]{1978ApJ...224..885W}
{Wilson}, R.~E. 1978, \apj, 224, 885, \dodoi{10.1086/156438}

\bibitem[{{Y{\i}ld{\i}z}(2014)}]{2014MNRAS.437..185Y}
{Y{\i}ld{\i}z}, M. 2014, \mnras, 437, 185, \dodoi{10.1093/mnras/stt1874}

\bibitem[{{Yildiz} \& {Do{\u{g}}an}(2013)}]{2013MNRAS.430.2029Y}
{Yildiz}, M., \& {Do{\u{g}}an}, T. 2013, \mnras, 430, 2029, \dodoi{10.1093/mnras/stt028}

\bibitem[{{Yu} {et~al.}(2022){Yu}, {Li}, {Huang}, {Hu}, \& {Xiang}}]{2022NewA...9101695Y}
{Yu}, Y.-X., {Li}, Q., {Huang}, H.-P., {Hu}, K., \& {Xiang}, F.-Y. 2022, \na, 91, 101695, \dodoi{10.1016/j.newast.2021.101695}

\bibitem[{{Zhang} {et~al.}(2020){Zhang}, {Qian}, \& {Liao}}]{2020MNRAS.492.4112Z}
{Zhang}, X.-D., {Qian}, S.-B., \& {Liao}, W.-P. 2020, \mnras, 492, 4112, \dodoi{10.1093/mnras/staa079}

\bibitem[{{Zhou} {et~al.}(2016{\natexlab{a}}){Zhou}, {Qian}, {Essam}, {He}, \& {Zhang}}]{2016NewA...47....3Z}
{Zhou}, X., {Qian}, S.~B., {Essam}, A., {He}, J.~J., \& {Zhang}, B. 2016{\natexlab{a}}, \na, 47, 3, \dodoi{10.1016/j.newast.2015.12.012}

\bibitem[{{Zhou} {et~al.}(2016{\natexlab{b}}){Zhou}, {Qian}, {Zhang}, {Zhang}, \& {Kreiner}}]{2016AJ....151...67Z}
{Zhou}, X., {Qian}, S.~B., {Zhang}, J., {Zhang}, B., \& {Kreiner}, J. 2016{\natexlab{b}}, \aj, 151, 67, \dodoi{10.3847/0004-6256/151/3/67}

\bibitem[{{Zhu} {et~al.}(2016){Zhu}, {Zhao}, \& {Zhou}}]{2016RAA....16...68Z}
{Zhu}, L.-Y., {Zhao}, E.-G., \& {Zhou}, X. 2016, Research in Astronomy and Astrophysics, 16, 68, \dodoi{10.1088/1674-4527/16/4/068}

\bibitem[{{Zwitter} {et~al.}(2003){Zwitter}, {Munari}, {Marrese}, {Pr{\v{s}}a}, {Milone}, {Boschi}, {Tomov}, \& {Siviero}}]{2003A&A...404..333Z}
{Zwitter}, T., {Munari}, U., {Marrese}, P.~M., {et~al.} 2003, \aap, 404, 333, \dodoi{10.1051/0004-6361:20030446}

\end{thebibliography}
\bibliographystyle{aasjournal}
\appendix  % 开始附录部分
% \clearpage % 清理浮动队列
% \appendix
% \section{Appendix}
% \clearpage
% \appendix

\renewcommand{\thetable}{A\arabic{table}}
\renewcommand{\thefigure}{A\arabic{figure}}
\setcounter{table}{0}
\setcounter{figure}{0}

\begin{figure*}[b]
	\begin{minipage}{0.4999\textwidth}
		\hspace{0.5pt}
        %这个图片路径替换成你的图片路径即可使用
		\centerline{\includegraphics[width=\linewidth]{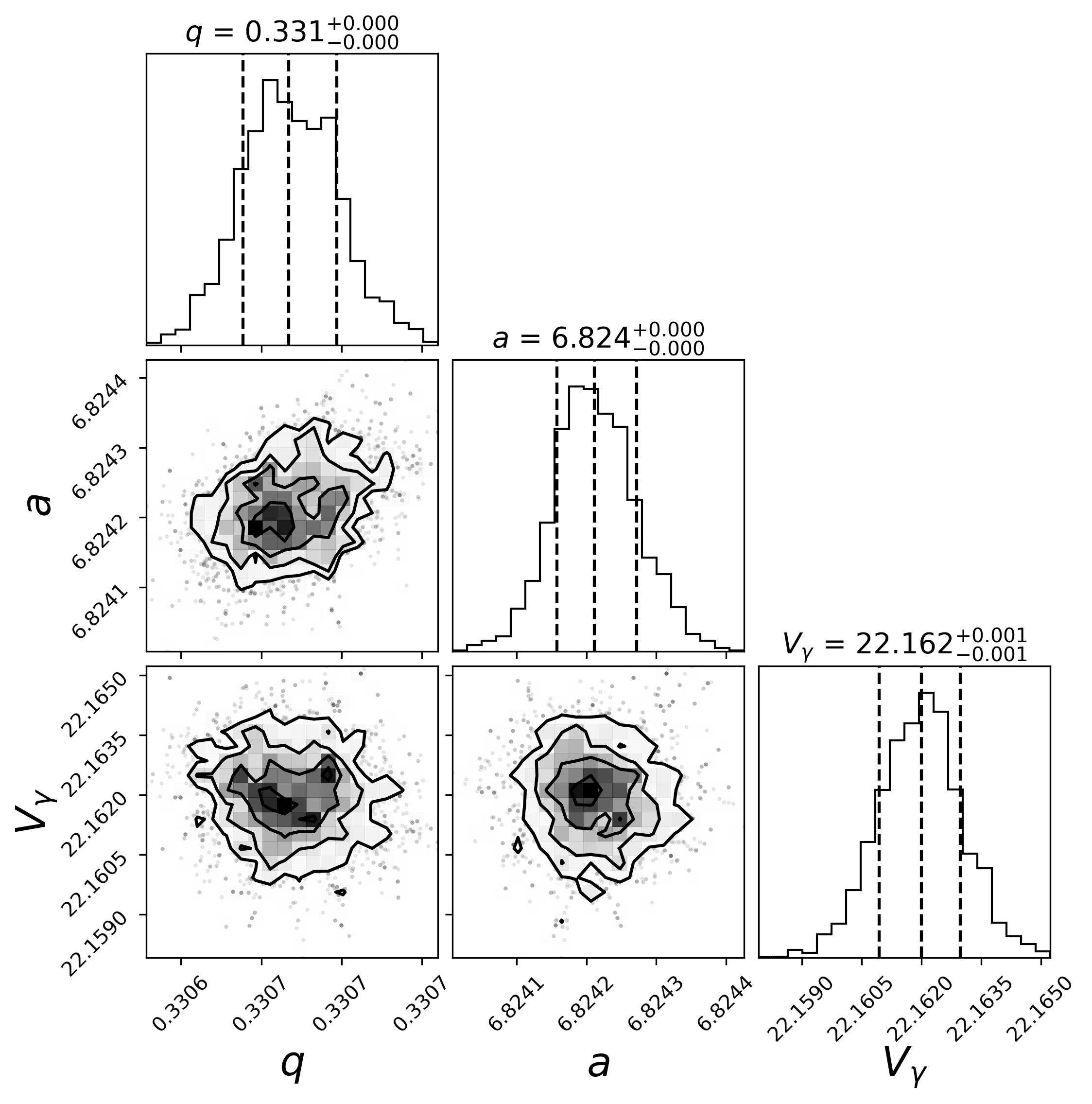}}
          % 加入对这列的图片说明
		%\centerline{Image}
	\end{minipage}
	\begin{minipage}{0.4799\textwidth}
		\hspace{0.5pt}
		\centerline{\includegraphics[width=\linewidth]{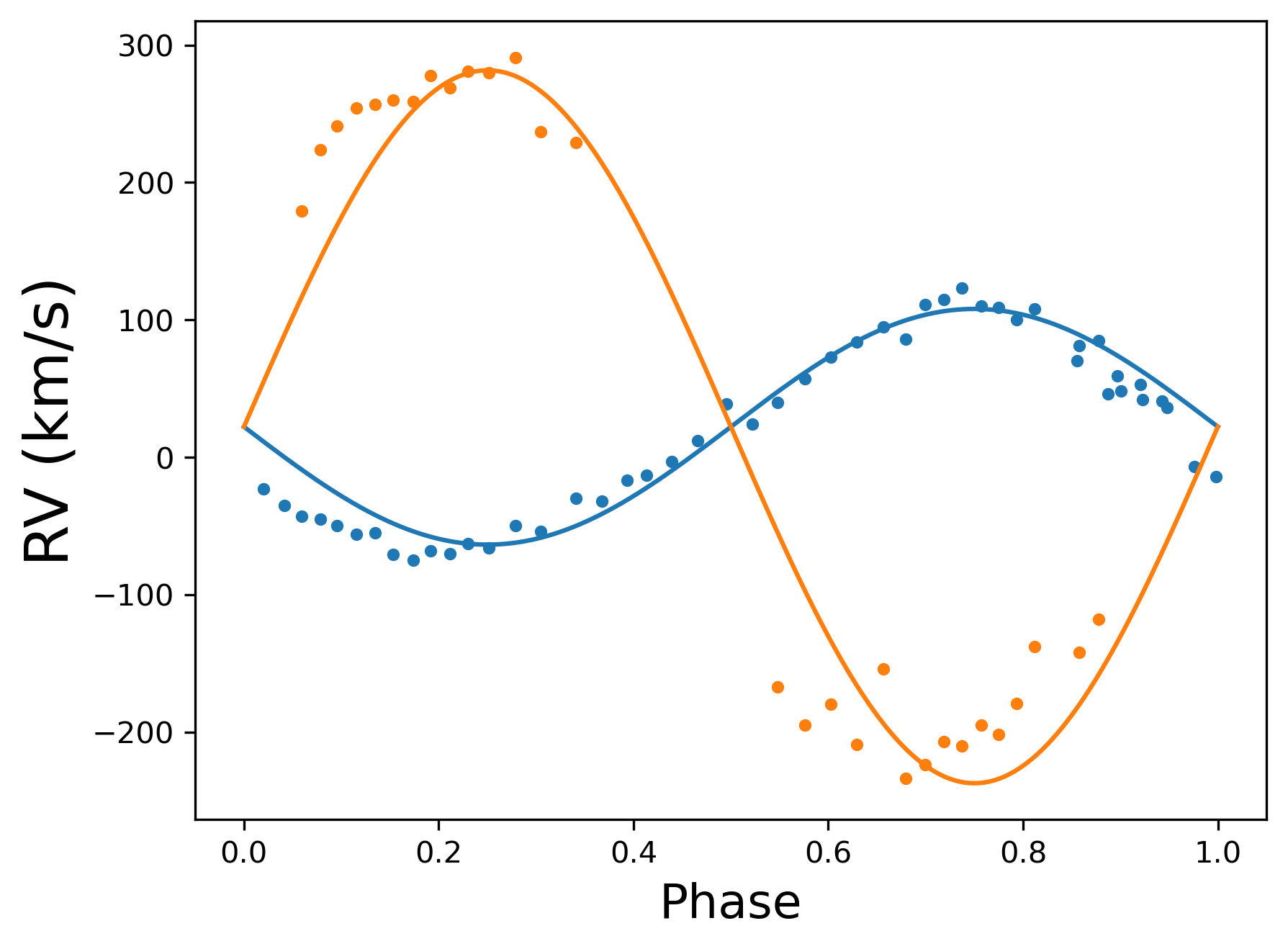}}
		%\centerline{Image}
	\end{minipage}
	\caption{The possible posterior distributions of $q$, $a$, $V_{\gamma}$ and fitting curve of radial velocity of TIC 267043786. $V_{\gamma}$ represents the radial velocity of the system.}
	\label{267043786}
\end{figure*}

\begin{figure*}
	\begin{minipage}{0.4999\textwidth}
		\hspace{0.5pt}
        %这个图片路径替换成你的图片路径即可使用
		\centerline{\includegraphics[width=\linewidth]{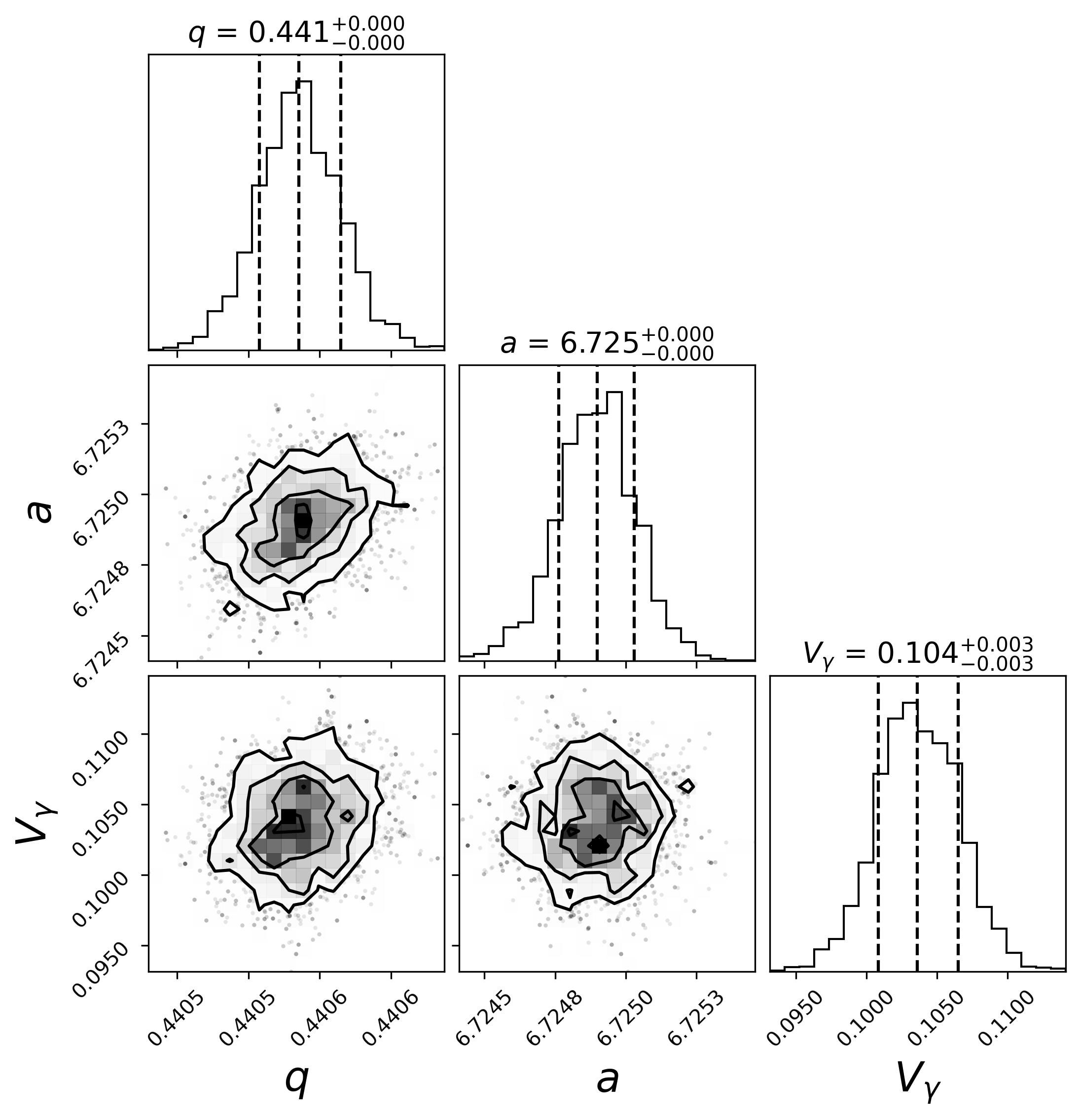}}
          % 加入对这列的图片说明
		%\centerline{Image}
	\end{minipage}
	\begin{minipage}{0.4799\textwidth}
		\hspace{0.5pt}
		\centerline{\includegraphics[width=\linewidth]{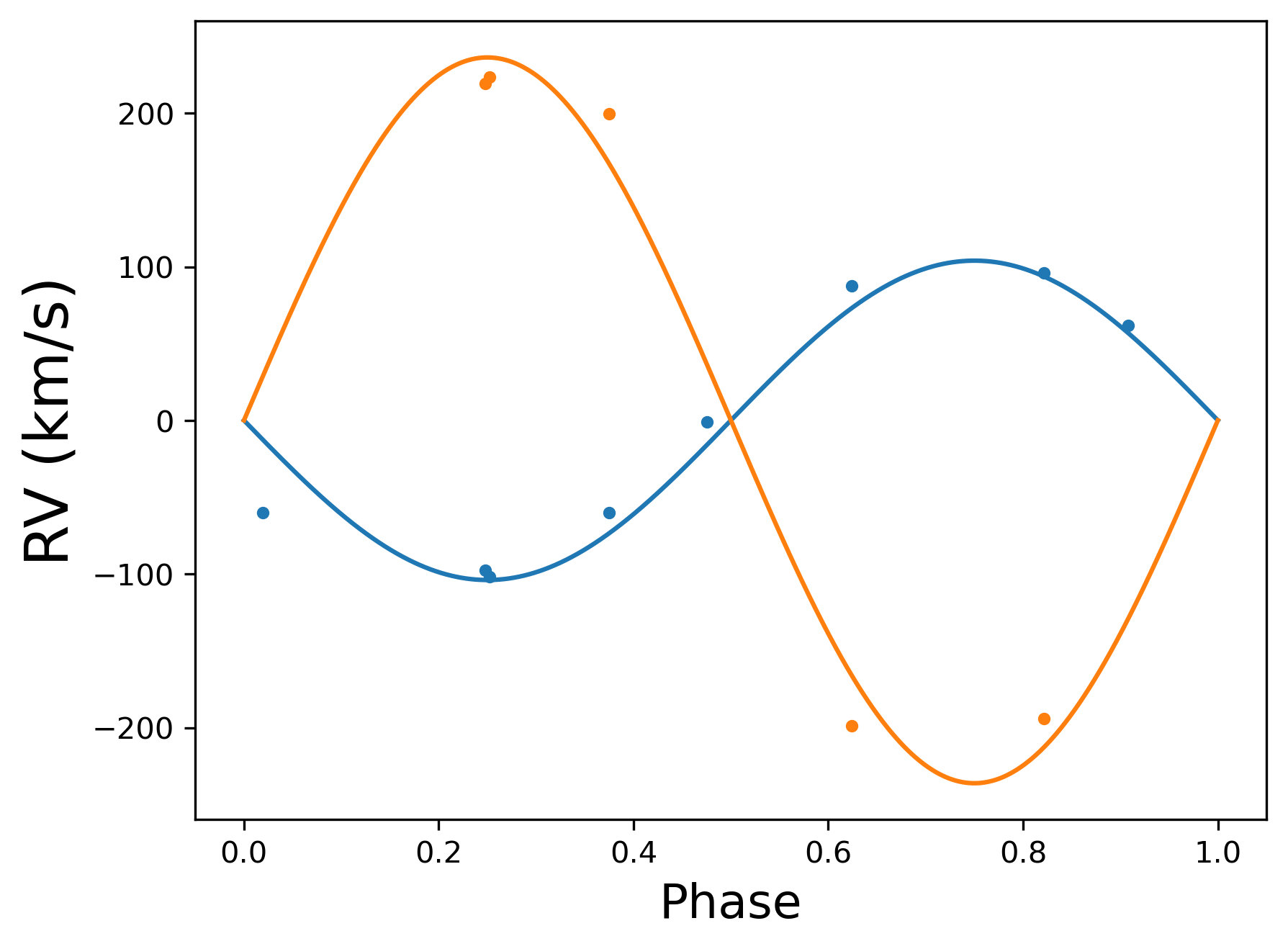}}
		%\centerline{Image}
	\end{minipage}
	\caption{The possible posterior distributions of $q$, $a$, $V_{\gamma}$ and fitting curve of radial velocity of TIC 207174531.}
	\label{207174531}
\end{figure*}

\begin{table}
\caption{Spot information of 21 targets.}
\centering 
\label{tab:spot}
\begin{tabular}{ccccccccc}
\hline
\hline
\multicolumn{1}{c}{\multirow{2}*{Targets(TIC)} }  & \multicolumn{3}{c}{Spot(star1)} &  \multicolumn{3}{c}{Spot(star2)} \\
%\cline{2-9}
\cmidrule(r){2-4} \cmidrule(r){5-7}
& $\lambda(deg)$   & $r_{s}(deg)$ &   $T_{s}$  &  $\lambda(deg)$   & $r_{s}(deg)$ &   $T_{s}$) \\
\hline 
101070833  &   $124_{-1}^{+1}$ &  $20_{-1}^{+1}$ &  $0.949_{-0.007}^{+0.007}$ &  …     &  …     &  …  \\
114500611  &   $1_{-1}^{+1}$ &  $22_{-1}^{+1}$ &  $0.925_{-0.012}^{+0.010}$ &     …  &     …  &     …   \\
117978580  &   $150_{0}^{+0}$ &  $27_{-3}^{+3}$ &  $0.954_{-1.058}^{+0.955}$ &   …    &   …    &   …   \\
124205037  &   $316_{-1}^{+1}$ &  $16_{0}^{+0}$ &  $0.88_{-0.017}^{+0.021}$ & …     & …     & …                                                      \\
147083089  &   $60_{-1}^{+2}$ &  $12_{0}^{+0}$ &  $0.879_{-0.022}^{+0.031}$ & …     & …     & …                                                      \\
159939533  &   $276_{-1}^{+1}$ &  $35_{0}^{+0}$ &  $0.936_{-0.019}^{+0.019}$ & …     & …     & …                                                     \\
164720673  &   $101_{-1}^{+1}$ &  $14_{-1}^{+1}$ &  $0.942_{-0.011}^{+0.010}$ &   …    &   …    &   …   \\
193580427  &   $2_{-1}^{+1}$ &  $33_{-1}^{+1}$ &  $0.936_{-0.006}^{+0.006}$ &    …   &    …   &    …   \\
197735761  &   $357_{-1}^{+1}$ &  $21_{-1}^{+1}$ &  $0.88_{-0.019}^{+0.019}$ &   …    &   …    &   …   \\
206537272  &   $68_{-1}^{+1}$ &  $14_{0}^{+0}$ &  $0.706_{-0.023}^{+0.024}$ & …     & …     & …                                                      \\
207174531  &   $99_{-1}^{+1}$ &  $20_{-1}^{+1}$ &  $0.964_{-0.006}^{+0.005}$ &   …    &   …    &   …   \\
232063593  &   $90_{-1}^{+1}$ &  $7_{-1}^{+1}$ &  $0.864_{-0.042}^{+0.037}$ &    …   &    …   &    …   \\
233233639  &   $4_{-1}^{+1}$ &  $28_{-1}^{+1}$ &  $0.952_{-0.006}^{+0.005}$ &    …   &    …   &    …   \\
241393039  &   $240_{-1}^{+1}$ &  $7_{0}^{+0}$ &  $0.745_{-0.023}^{+0.023}$ & …     & …     & …                                                      \\
244207973  &   $88_{-1}^{+1}$ &  $17_{-1}^{+1}$ &  $0.86_{-0.018}^{+0.018}$ &    …   &    …   &    …   \\
254298146  &   $3_{-1}^{+1}$ &  $57_{-1}^{+1}$ &  $0.957_{-0.002}^{+0.002}$ &    …   &    …   &    …   \\
255586667  &   $12_{-1}^{+1}$ &  $46_{-1}^{+1}$ &  $0.954_{-0.003}^{+0.003}$ &   …    &   …    &   …   \\
261089147  &   $82_{-1}^{+1}$ &  $35_{-1}^{+1}$ &  $0.953_{-0.003}^{+0.003}$ &   …    &   …    &   …   \\
267043786  &   $174_{-1}^{+1}$ &  $10_{-1}^{+1}$ &  $0.611_{-0.034}^{+0.037}$ &  …     &  …     &  …   \\
269505148  &   $14_{-1}^{+1}$ &  $42_{-1}^{+1}$ &  $0.948_{-0.003}^{+0.002}$ &   …    &   …    &   …   \\
279481568  &   $92_{-1}^{+2}$ &  $45_{-1}^{+1}$ &  $0.953_{-0.004}^{+0.005}$ &   …    &   …    &   …   \\
290034099  &   $126_{-1}^{+1}$ &  $31_{-1}^{+1}$ &  $0.948_{-0.004}^{+0.004}$ &  …     &  …     &  …   \\
314459000  &   $358_{-2}^{+3}$ &  $15_{-2}^{+2}$ &  $1.046_{-0.030}^{+0.063}$ &   …    &   …    &   …   \\
316250867  &   $99_{-1}^{+1}$ &  $14_{-1}^{+1}$ &  $0.741_{-0.030}^{+0.035}$ &    …   &    …   &    …   \\
32284829   &   $305_{-1}^{+1}$ &  $23_{-1}^{+1}$ &  $0.824_{-0.014}^{+0.013}$ &  $195_{-1}^{+1}$ &  $34_{-1}^{+1}$ &  $0.747_{-0.035}^{+0.024}$   \\
332910986  &   $127_{-4}^{+4}$ &  $18_{-1}^{+1}$ &  $0.929_{-0.008}^{+0.008}$ & …     & …     &  …                         \\
347552724  &   $78_{-1}^{+1}$ &  $15_{-1}^{+1}$ &  $0.794_{-0.030}^{+0.029}$ &  $175_{-1}^{+1}$ &  $38_{-1}^{+1}$ &  $0.807_{-0.022}^{+0.025}$     \\
356192212  &   $85_{-1}^{+1}$ &  $25_{-1}^{+1}$ &  $0.947_{-0.005}^{+0.005}$ & …     & …     & …                                                     \\
383722507  &   $272_{-1}^{+1}$ &  $8_{-1}^{+1}$ &  $0.709_{-0.032}^{+0.048}$ & …     & …     & …                                                     \\
387586260  &   $91_{-1}^{+1}$ &  $13_{-1}^{+1}$ &  $0.594_{-0.052}^{+0.035}$ & …     & …     & …                                                     \\
393978805  &   $186_{-1}^{+2}$ &  $6_{-1}^{+1}$ &  $1.04_{-0.086}^{+0.045}$ & …     & …     & …                                                      \\
399577123  &   $248_{-3}^{+2}$ &  $21_{-1}^{+1}$ &  $0.921_{-0.009}^{+0.007}$ & …     & …     & …                                                    \\
399788405  &   $3_{-1}^{+1}$ &  $26_{-1}^{+1}$ &  $0.906_{-0.009}^{+0.009}$ & …     & …     & …                                                      \\
401928563  &   $75_{-1}^{+1}$ &  $8_{-1}^{+1}$ &  $0.885_{-0.029}^{+0.026}$ & …     & …     & …                                                      \\
402324760  &   $268_{0}^{+0}$ &  $16_{0}^{+0}$ &  $0.735_{-0.028}^{+0.024}$ & …     & …     & …                                                      \\
412063998  &   $336_{-2}^{+3}$ &  $9_{0}^{+0}$ &  $0.823_{-0.213}^{+0.037}$ & …     & …     & …                                                      \\
422914426  &   $73_{-1}^{+1}$ &  $31_{-1}^{+1}$ &  $0.954_{-0.003}^{+0.003}$ & …     & …     & …                                                     \\
455064763  &   $304_{-1}^{+1}$ &  $14_{0}^{+0}$ &  $0.818_{-0.021}^{+0.019}$ & …     & …     & …                                                     \\
53842685   &   $102_{-5}^{+4}$ &  $5_{-1}^{+1}$ &  $0.722_{-0.044}^{+0.043}$ & …     & …     & …                                                     \\
67686223   &   $358_{-1}^{+1}$ &  $14_{-1}^{+1}$ &  $0.761_{-0.022}^{+0.027}$ &  $215_{-1}^{+1}$ &  $24_{-1}^{+1}$ &  $0.808_{-0.025}^{+0.026}$   \\
71107089   &   $134_{-4}^{+8}$ &  $28_{-1}^{+1}$ &  $0.965_{-0.003}^{+0.003}$ & …     & …     & …                                                    \\
75377304   &   $258_{-1}^{+1}$ &  $15_{-1}^{+1}$ &  $0.871_{-0.024}^{+0.017}$ & …     & …     & …                                                    \\
8989778    &   $91_{-1}^{+1}$ &  $27_{-1}^{+1}$ &  $0.955_{-0.004}^{+0.004}$ & …     & …     & …                                                     \\
\hline
\end{tabular}
\end{table}

%% This command is needed to show the entire author+affiliation list when
%% the collaboration and author truncation commands are used.  It has to
%% go at the end of the manuscript.
%\allauthors

%% Include this line if you are using the \added, \replaced, \deleted
%% commands to see a summary list of all changes at the end of the article.
%\listofchanges

\end{document}